\documentclass[usenatbib]{mnras}

\usepackage{newtxtext,newtxmath}

\usepackage{color}

\newcommand{\Msun}{{\ M_{\odot}} }

\usepackage[T1]{fontenc}
\usepackage{ae,aecompl}

\usepackage{graphicx}	
\usepackage{amsmath}	
\usepackage{amssymb}	
\usepackage{float}
\usepackage{longtable}
\usepackage{wasysym}
\usepackage{hyperref}

\graphicspath{{./}{Images/}}

\title{Low Mass X-ray Binaries: The Effects of the Magnetic Braking Prescription}

\author[Kenny Van]{
K.~X. Van$^{1}$\thanks{E-mail: kvan@ualberta.ca}, N. Ivanova$^{1}$, C.~O. Heinke$^{1}$
\\
$^{1}$University of Alberta, Dept. of Physics, 11322-89 Ave, T6G 2E7, Canada\\
}

\date{Accepted XXX. Received YYY; in original form ZZZ}

\pubyear{2018}

\begin{document}

\label{firstpage}
\pagerange{\pageref{firstpage}--\pageref{lastpage}}
\maketitle
\begin{abstract}

We present a population study of low- and intermediate-mass X-ray binaries (LMXBs) with neutron star accretors, performed using the detailed 1D stellar evolution code {\tt MESA}. 
We identify all plausible Roche-lobe overflowing binaries at the start of mass transfer, and compare our theoretical  mass-transfer tracks to the population of well-studied Milky Way LMXBs. The mass-transferring evolution depends on the accepted magnetic braking (MB) law for angular momentum loss. The most common MB prescription ("Skumanich MB") originated from observations of the time-dependence of rotational braking of Sun-type stars, 
where the angular momentum loss rate depends on the donor mass $M_d$, donor radius $R_d$, and rotation rate $\Omega$, $\dot{J} \propto M_d R_d^{\gamma} \Omega^3$. 
The functional form of the Skumanich MB can be also obtained theoretically assuming a radial magnetic field, isotropic isothermal winds, and boosting of the magnetic field by rotation. Here we show that this simple form of the Skumanich MB law 
gives mass transfer rates an order of magnitude too weak to explain most observed persistent LMXBs. This failure suggests 
that the standard Skumanich MB law should not be employed to interpret Galactic, or extragalactic, LMXB populations, 
with either detailed stellar codes or rapid binary population synthesis codes. We investigate modifications for the MB law, 
and find that including a scaling of the magnetic field strength with the convective turnover time, and a scaling of MB with the wind mass loss rate, can reproduce persistent LMXBs, and does a better job at reproducing transient LMXBs.

\end{abstract}

\begin{keywords}
methods: numerical -- binaries: general -- stars: magnetic field -- stars: evolution -- X-rays: binaries.
\end{keywords}



\section{Introduction}
\label{Introduction}

Low mass X-ray binaries (LMXBs) with a neutron star (NS) accretor are among the most well-studied binary systems in astrophysics. Over 100 of them have been observationally identified in the Milky Way over the last 50 years, while a plethora of binary parameters -- mass ratios, orbital periods, mass transfer (MT) rates, and donors' effective temperatures in some cases -- have been determined for several of them. For many years, X-ray binaries have posed as an enticing problem for theorists, providing grounds to verify ways to model binary stellar evolution \citep{Faulkner1971, Rappaport1983, Webbink1983, Joss1984, Podsiadlowski2002}. The applications of theoretical models of LMXBs are not limited to only our Galaxy, but have also been  applied to interpret the observed populations of X-ray binaries in other galaxies \citep[e.g.,][]{Fragos2008, Tremmel2013}, and even to estimate their feedback on reionization of the early universe \citep[e.g.,][]{Fragos2013}

Understanding and interpreting LMXBs can be split into two distinct stages: the formation and life of a binary system with a NS before the start of the MT, and the evolution of the binary system during the MT. Before the onset of the MT, the evolution of the binary is expected to proceed through a a common envelope phase and a supernova explosion   \citep{Bhattacharya1991}. Supernova natal kicks, supernova explosion mechanisms, and widely used pulsar kick distributions, while observationally derived, are not fully understood \citep{Fryer2012, Hobbs2005}. The outcomes of common envelope events are highly uncertain as well \citep[for a review, see][]{Ivanova2013b}.

In this paper, we focus on the evolution of LMXBs during MT. The driving mechanism of the MT phase is the donor's response to the mass and angular momentum loss. In short period systems, the dominant channel of angular momentum loss is gravitational radiation, which is well understood theoretically and has been confirmed by observations \citep{Weisberg2016, Abbott2016, Abbott2017}. In cases with longer orbital periods, magnetic braking (MB) is the dominant angular momentum loss mechanism \citep{Rappaport1983}.The strength of MB affects the evolution of LMXBs by increasing or decreasing the angular momentum loss of the binary. In cases where the angular momentum loss is greater, the binary will tend to shrink and thus undergo more aggressive MT. The default assumption for MB is to adopt the "Skumanich" empirical law where the angular momentum loss scales with donor mass $M_d$, donor radius $R_d$, and rotation rate $\Omega$, $\dot{J} \propto M_d R_d^\gamma \Omega^3$ \citep{Skumanich1972}. In theoretical models, this empirical law is used by employing the parameterized prescriptions for MB stated by either \cite{Verbunt1981} or \cite{Rappaport1983}. Numerical studies of LMXB populations have shown systematic mismatches between simulated results and observations. Observed MT rates have been found differing from the predicted rates in simulated systems by up to an order of magnitude \citep{Podsiadlowski2002}. Despite obtaining more detailed data on LMXBs, the discrepancies remained prevalent. For instance, Sco X-1 cannot be explained by the "Skumanich" law.  To match the observations of Sco X-1, the rate of angular momentum loss has to be boosted, for example by taking into account the effect of stellar wind loss \cite{Pavlovskii2015}.

We are taking this analysis a step further by including the effects of the convective turnover time, as well as considering a non-isothermal stellar wind \S~\ref{sec:mb}. We apply different MB laws to a grid of 2136 binary systems spanning reasonable initial conditions, for a circularized binary system with a NS, where MT starts within 10 Gyr \S~\ref{sec:Binary_calculations}. We describe the principal properties of the simulated MT systems in \S~\ref{sec:Principle Results} and have compiled a table with  updated properties of some observed NS LMXBs (see \S~\ref{sec:Observational Data}). The observational data is then used to constrain the MB laws in \S~\ref{sec:stat}.

\section{Numerical Method}

\subsection{Setting up and evolving the population of binaries}

\label{sec:Binary_calculations}

We consider the set of initial binary systems as follows:

\begin{itemize}
\item The masses of donors range from $M_{\rm d}=1.0\Msun$ to $M_{\rm d}=7 \Msun$. The grid in donor masses uses steps of $1 \Msun$ for donors with masses  $\ge 5 \Msun$, $0.5 \Msun$ for donors between $3 \Msun$ and $5 \Msun$, 0.2 $\Msun$ for donors between $2.4 \Msun$ and $3 \Msun$, and $0.1 \Msun$ for donors with masses $\le 2.4 \Msun$. This is the mass of the donors at ZAMS. The adopted metallicity is $Z=0.02$. 

\item The initial binary orbital periods range from the periods at which the donor stars would overflow their Roche lobes at ZAMS, to the maximum orbital period at which they would start the MT while they are on the red giant branch. We define the initial period as the period at ZAMS of the donor star; which is not the same as the period that a binary would have at the start of the MT. There can be a large difference between when the initial period at the donor stars ZAMS and at the onset of RLOF. The mesh for the initial orbital periods has a fixed step of 0.05 in $\log_{10}(P)$.  All orbits are circular.

\item With the initial mesh of seed masses and periods, we ran 2136 simulations for each considered MB scheme (see \S~\ref{sec:mb}).

\item The compact companions are NSs with mass $M_{\rm a}=1.4 \Msun$ and radius $R_{\rm a}=11.5$~km.
  
\end{itemize}

All calculations for the single stars and mass transferring binaries were performed using the one-dimensional stellar evolution code \texttt{MESA} \footnote{\url{http://mesa.sourceforge.net}} (Modules for Experiments in Stellar Astrophysics), revision 10398 and the August 2018 release of the \texttt{MESASDK} \footnote{\url{http://www.astro.wisc.edu/~townsend/static.php?ref=mesasdk}}. \texttt{MESA} is a modern open-source set of stellar libraries as described in \cite{Paxton2011, Paxton2013, Paxton2015, Paxton2018}. The donors are evolved using default assumptions as in \texttt{MESA}. In particular, we use the mixing length $\alpha=2$, no semiconvection, and no overshooting. {\tt MESA} uses mixing length theory as described in \cite{Cox1968}. We adopt a grey atmosphere boundary condition and use the OPAL opacity tables for  solar composition \citep{Grevesse1993}\footnote{The inlists files can be found at the \texttt{MESA} marketplace website: \url{http://cococubed.asu.edu/mesa_market/}}.

The donor stars are evolved using Reimer's wind mass loss prescription \citep{Reimer1975}: 

\begin{equation}
    \dot{M}_{\rm d}^{\rm w} = \eta \times 4 \times 10^{-13} \frac{R}{R_\odot}\frac{L}{L_\odot}\frac{M_\odot}{M} M_\odot \rm{yr^{-1}} ,
    \\
    \eta = 1
\end{equation}

\noindent
$\eta$ is a scaling factor or efficiency of wind loss. In our calculations we use $\eta=1$.




When the donor star overfills its Roche lobe we calculate MT; we do not consider any other forms of MT. For the Roche lobe radius, we use the fitting formula for the volume equivalent one-dimensional star's radius, as provided by \cite{Eggleton1983}:

\begin{equation}
    \frac{R_L}{a}=\frac{0.49q^{2/3}}{0.6q^{2/3}+\ln(1+q^{1/3})} \ .
\end{equation}

\noindent
Here $R_L$ is the Roche lobe radius of the donor star with mass $M_{\rm d}$, $a$ is the orbital separation between the two stars, and $q=M_{\rm d}/M_{\rm a}$ is the mass ratio of the two stars. To calculate the MT via the Lagrange point between the two stars $L_1$, $\dot M_{\rm d}^{L1}$, we employ the "Ritter" mass loss scheme, as implemented in \texttt{MESA} \citep[see][for details of this MT prescription]{Ritter1988}. We consider that the MT may be non conservative. If $\dot M_{\rm d}^{L1}$ exceeds the Eddington-limited MT rate $\dot{M}_{\rm Edd }$, the excess $\dot M_{\rm d}^{L1}-\dot{M}_{\rm Edd }$ cannot be accreted on to the compact object, and is assumed to be lost from the system. For the Eddington limited MT rate on a NS we use, 

\begin{equation}
    \dot{M}_{\rm Edd }=\frac{4\pi c R_{\rm a}}{\kappa_{e}} \approx \frac{3.4}{1+X}\times 10^{-8} M_\odot \rm {\ yr}^{-1} 
    \label{eq:MT_Edd}
\end{equation}

\noindent Here $\kappa_e$ is the opacity due to Thomson electron scattering, $\kappa_{e}=0.2(X+1)$ cm$^2$ g$^{-1}$, where X is the hydrogen mass fraction in the material transferred from the donor.

The angular momentum of the system is lost through gravitational radiation, or through MB, or is carried away with the mass lost from the system. The mass lost from the donor due to wind mass loss leaves with the specific angular momentum of the donor. Note the orbital evolution is calculated before the start of the MT as well. If the MT rate exceeds $\dot{M}_{\rm Edd }$, the material exceeding the Eddington limit is lost with the specific angular momentum of the accretor. The angular momentum loss due to gravitational radiation is found using the standard formula \citep{Faulkner1971}:

\begin{equation}
\frac{\dot{J}_{\rm{GR}}}{J} = -\frac{32}{5} G^3 c^5 \frac{M_{\rm d} M_{\rm a} (M_{\rm d} + M_{\rm a})}{a^4}
\end{equation}

\subsection{Magnetic braking}

\label{sec:mb}

We calculate the MT sequences considering several prescriptions for the angular momentum loss via MB. The first prescription considered uses the formulation from \cite{Rappaport1983}:

\begin{equation}
\dot{J}_{\rm{MB,Sk}} = -3.8 \times 10^{-30} M_{\rm d}R_{\odot} ^4 \left(\frac{R_{\rm d}}{R_\odot}\right)^{\gamma_{\rm mb}} \ \Omega^3 \rm{dyne} \ \rm{cm}
\label{eq:def_MB}
\end{equation}

\noindent Here, $R_{\rm d}$ is the radius of the donor, $\gamma_{\rm mb}$ is a dimensionless parameter from 0 to 4 and $\Omega$ is the angular velocity of the donor. The stars are kept in corotation with the binary as the angular velocity is calculated using the binary period. With $\gamma_{\rm mb}=4$,  Equation~\ref{eq:def_MB} describes the standard Skumanich law 
as derived by \citet{Verbunt1981}, 
and is the most commonly used form for MB in calculations of LMXB evolution. 

The Skumanich law was scaled to describe main sequence stars similar to our Sun. In systems with donors different from the Sun, the increased rate of mass loss with the stellar wind, 
as well as magnetic field strength that does not scale directly with the angular velocity of the star, will play a role in the MB calculation \citep{Mestel1968, Mestel1987, Kawaler1988}. To determine the effects of these additional terms, we follow similar steps as \cite{Pavlovskii2015}. We start with the formulation given by \cite{Mestel1987} which parameterizes the amount of angular momentum lost from the system through mass leaving through the Alfv\`en surface. The Alfv\`en surface is the surface where the ram pressure is equal to the magnetic pressure \citep{Mestel1968, Mestel1987}:

\begin{equation}
    \frac{1}{2} \rho_{\rm A} v_{\rm A}^2 \simeq \frac{B(r)^2}{8 \pi}.
\label{eq:alfven}
\end{equation}

$\rho_{\rm A}$ and $v_{\rm A}$ denote the density and velocity of the wind as it crosses the Alfv\`en surface. $B(r)$ is the poloidal magnetic field strength as a function of radius. This value encompasses the structure of the magnetic field of the star. In the simplest case where the magnetic field is radial, and $B_s$ is the surface magnetic field strength, then:

\begin{equation}
    B(r) = B_s \frac{R_s^2}{r^2}.
\label{eq:radial_b}
\end{equation}

Throughout this work, we will only be using a radial magnetic field, but it should be noted that other magnetic field structures are possible. In the context of magnetic braking however, the Alfv\`en surface represents the maximum radius where the stellar wind is locked in corotation with the surface of the star, beyond this point the mass is assumed to be lost \citep{Mestel1968, Mestel1987}. The equation which describes the angular momentum loss through an Alfv\`en surface is:


\begin{equation}
\begin{split}
    \dot{J}_{\rm MB} &= - 4\pi \Omega \int_0^{\pi/2} \rho_{\rm A} v_{\rm A} R^2_{\rm A} (R_{\rm A} \sin{\theta})^2 \sin{\theta} d\theta \\
                     &\simeq -\frac{8}{3} \pi \Omega \rho_{\rm A} v_{\rm A} R^4_{\rm A}, \\
                     &\simeq -\frac{2}{3} \Omega \dot{M}_{\rm W} R^2_{\rm A}.
\label{eq:mb_init}
\end{split}
\end{equation}

Here it is assumed that the Alfv\`en surface, $R_{\rm A}$ does not depend on $\theta$, the polar angle, and that the wind coming from the star is isotropic. Should the system be rapidly rotating the scaling of the magnetic braking to the rotation rate will change as so-called dead zones may form \citep[see discussion in][]{Ivanova2006}.  It is important to note the \cite{Mestel1987} parameterization can account for ``dead zones'' where material is trapped in magnetic fields lines and not lost through the Alfv\`en surface. The material confined within the dead zone remain in corotation within the dipole field and is not lost from the system. We don't include the effects of dead zones in this work and as such may over estimate the amount of material lost and thus the angular momentum loss in tight binaries. The wind mass loss rate, if isotropic can be described using the following equation:

\begin{equation}
\begin{split}
    \dot{M}_{\rm W} &= 4 \pi R^2 \rho_s v_s, \\ 
                    &= 4 \pi R^2_{\rm A} \rho_{\rm A} v_{\rm A},
\label{eq:stellar_wind}
\end{split}
\end{equation}

\noindent Combining equations \ref{eq:alfven} and \ref{eq:radial_b} gives us:

\begin{equation}
    4 \pi R_A^4 \rho_A v_A^2 = B_s^2 R^4,
\label{eq:alfven_mod}
\end{equation}

\noindent Including the wind mass loss equation as given in equation \ref{eq:stellar_wind}, equation \ref{eq:alfven_mod} becomes:

\begin{equation}
\begin{split}
    \dot{M}_{\rm W} R_A^2 v_A &= B_s^2 R^4 ,\\
    R_A^2 &= \frac{B_s^2 R^4}{\dot{M}_{\rm W} v_A} .
\label{eq:alfven_rad_polodial}
\end{split}
\end{equation}

Under the assumption of isothermal winds, the wind reaches a sonic wind velocity at the Alfv\`en surface of $v_{\rm A} = c_w$ \citep{Mestel1987} where $c_w$ is a constant value. Combining equations \ref{eq:mb_init} and \ref{eq:alfven_rad_polodial} gives the following MB scaling equation: 

\begin{equation}
    \dot{J}_{\rm MB} \propto \Omega B_s^2 R^4.
    \label{eq:sku_jdot}
\end{equation}

Equation~\ref{eq:sku_jdot} interestingly, does not contain any scaling with the stellar wind, despite the wind strength being a fundamental physical property of MB. The assumption of isothermal winds does not apply to giant stars. In the case of giant stars with cooler temperatures, winds may be accelerated by a variety of mechanisms, this requires a different self-consistent description of the wind velocity \citep{Suzuki2007}. In the case of a radial field and a nonthermal wind where the wind moves at speeds on order of the escape velocity,

\begin{equation}
    v_{\rm A}^2 = \frac{2GM}{R},
\end{equation}

\noindent We get the following scaling:

\begin{equation}
    \dot{J}_{\rm MB} \propto \dot{M}_{\rm W} \Omega B_s^4 R^4.
    \label{eq:radial_nontherm}
\end{equation}

\noindent The strength of the surface magnetic field, $B_s$, scales with the dynamo number $N_D$ \citep{Parker1971}. The dynamo number is related to physical values in the MB through the Rossby number, $R_0$ \citep{Noyes1984},

\begin{equation}
\begin{split}
    N_d & \approx R_0^{-2}, \\
    N_d & \approx \Omega^2 \tau^2_{\rm conv} .
\end{split}
\end{equation}

\noindent Here $\tau_{\rm conv}$ is the turnover time of convective eddies, 

\begin{equation}
\tau_{\rm conv} = \int_{R}^{R_{s}} \frac{dr}{v_{\rm conv}} .
\end{equation}

\noindent $R$ and $R_s$ are the bottom and the top of the outer convective zone respectively, while $v_{\rm conv}$ is the local convective velocity. We follow a simple approximation made by \cite{Ivanova2006} where $B_s \propto N_d^{1/2}$. This allows us to adopt the scaling relations:

\begin{equation}
\begin{split}
    \frac{B_s}{B_{s,\odot}} &= \frac{R_0}{R_{0, \odot}}, \\
    \frac{B_s}{B_{s,\odot}} &= \frac{\Omega}{\Omega_\odot} \frac{\tau_{\rm conv}}{\tau_{\odot \rm conv}} .
\end{split}
\label{eq:B_scaling}
\end{equation}

\noindent Work by \cite{Auriere2015} has shown a general correlation between the semi-empirical Rossby number and the observed magnetic field strength of a star.
\cite{Auriere2015} also noted that dwarf stars have a steeper relation between these properties than giant stars.
In general, dwarf stars have been found to have shorter rotation periods with strong magnetic fields. As such, the relation used likely underestimates the strength of the magnetic field in dwarf stars. Rewriting the magnetic field scaling in Equation~\ref{eq:sku_jdot} gives us the following:

\begin{equation}
\begin{split}
    \dot{J}_{\rm MB} &\propto \Omega B_s^2 R^4, \\
                     &\propto \Omega^3 \tau_{\rm conv}^2 R^4.
\end{split}
\label{eq:jdot_rad_isotherm}
\end{equation}

The radial isothermal approximation results in the Skumanich scaling with $\Omega^3 R^4$ if we ignore the convective turnover time $\tau_{\rm conv}$. The radial non-thermal approximation from Equation~\ref{eq:radial_nontherm} becomes

\begin{equation}
\begin{split}
    \dot{J}_{\rm MB} &\propto \dot{M}_{\rm W} \Omega B_s^4 R^4\\
                     &\propto \dot{M}_{\rm W} \Omega^5 \tau_{\rm conv}^4 R^4.
\end{split}
\label{eq:jdot_rad_nontherm}
\end{equation}

Rewriting the Skumanich law to include the additional terms for wind $\dot{M}_{\rm W}$, convective turnover time $\tau_{\rm conv}$ and rotation rate $\Omega$, the general MB equation we use will be

\begin{equation}
    \dot{J}_{\rm MB, boost} = \dot{J}_{\rm MB, Sk} \left(\frac{\Omega}{\Omega_\odot}\right)^\beta \left( \frac{\tau_{\rm conv}}{\tau_{\odot, \rm conv}} \right)^\xi \left(\frac{\dot{M}_{\rm W}}{\dot{M}_{\odot,W}} \right)^\alpha.
\label{eq:mod_MB}
\end{equation}

The value used to normalize the convective turnover time, $\tau_{\odot, \rm conv} = 2.8 \times 10^6 \ \rm s$, was found by evolving a $1 \Msun$ star at $Z$=0.020 to 4.6 Gyrs. Similarly, the solar wind value $\dot{M}_{\odot, \rm W} = 2.5\times10^{-14} M_\odot \rm \ yr^{-1}$ \citep{Carroll2006}, and $\Omega_\odot \approx 3\times10^{-6} \ \rm s^{-1}$ is the angular frequency of the Sun using an orbital period of 24 days.

The power $\xi$ can vary, where $\xi=0$ describes the same simplified assumptions for which the Skumanich law is valid with $\alpha=0$ and $\beta=0$ (i.e., radial magnetic field and isothermal winds). $\xi=2$ is the case described in equation \ref{eq:jdot_rad_isotherm} which results in the convection boosted Skumanich case. $\xi$ may be as high as 4 for the case of winds from giants where the velocity of the wind grows linearly with distance; we note that in this case, the dependence on the angular velocity will also have to be modified to $\Omega^5$, vs. the Skumanich law's factor of $\Omega^3$. Therefore in this case, $\xi=4$, $\beta=2$ and $\alpha=1$.






We will use these additional scaling terms and define different MB cases for the tested grid of binaries:

\begin{enumerate}

\item ``Default'': We use the default MB scheme described by \cite{Rappaport1983}, without the additions mentioned in \cite{Pavlovskii2015}.  $\gamma_{\rm mb} = 4$ in this case, and all subsequent cases.

\item ``Convection-boosted'': We adopt the scaling found in Equation~\ref{eq:jdot_rad_isotherm}, which is the Skumanich law,  scaled by the convective turnover time $(\tau_{\rm conv})^{\xi}$. The value of $\xi = 2$ will be used in this prescription.

\item ``Intermediate'': We use the convection-boosted MB scheme and apply an additional wind scaling term, linear in wind mass loss rate ($\alpha = 1$). 

\item ``Wind-boosted'': This MB scheme uses the scaling values from Equation~\ref{eq:jdot_rad_nontherm}. This prescription includes all three scaling terms shown in Equation~\ref{eq:mod_MB} with $\beta=2$, $\xi=4$ and $\alpha=1$. 

\end{enumerate}

\begin{table}
    \centering
    \begin{tabular}{l|ccc}
        Case                   & $\beta$ & $\xi$ & $\alpha$  \\
        \hline
        1 - Default Skumanich    & 0       & 0     & 0         \\
        2 - Convection Boosted  & 0       & 2     & 0         \\
        3 - Intermediate              & 0       & 2     & 1         \\
        4 - Wind Boosted            & 2       & 4     & 1         \\
    \end{tabular}
    \caption{The different scaling values used in Equation~\ref{eq:mod_MB} for the various cases.}
    \label{tab:MB_scalings}
\end{table}

These systems are evolved to 10 Gyrs, or until the donor star loses its envelope and detaches. If the simulation encounters dynamically unstable MT, which \texttt{MESA} was not designed to adequately model, the system will likely encounter numerical issues and stop. We do not consider irradiation effects on the companion star during its evolution (see \S 6).

\subsection{Verification against a previous study}
\label{subsec:ScoX1}

Of particular importance is the binary system Scorpius X-1 as this system allows us to compare our results to the work of  \cite{Pavlovskii2016}. In \cite{Pavlovskii2016}, the authors used a modified MB prescription where $\alpha = 1$, $\beta = 0$, and $\xi = 0$, with an MB gamma $\gamma_{\rm mb}$ of 3. In this work, we tested all MB prescriptions described in \S \ref{sec:Binary_calculations} and used an MB gamma $\gamma_{\rm mb}$ of 4. To ensure the changes to MB were correctly implemented, the comparisons between models used in \cite{Pavlovskii2016} were rerun.

Sco X-1 is an LMXB, observed to have a mass ratio in the range  from 0.28 - 0.51, favouring a value of $\approx$ 0.30 \citep{Sanchez2015, Steeghs2002}.  The NS is constrained to have a mass of $<$ 1.73 $\Msun$ \citep{Sanchez2015}. The period of Sco X-1 is 18.8951 hours, and the MT rate is estimated to be at least $\sim 2.2 \times 10^{-8}~\Msun yr^{-1}$ \citep{Watts2008, Pavlovskii2016}. Observations of this system provided upper limits on the spectral class of the donor of K4 or later, with the luminosity class IV, and the implied effective temperature  less than 4800 K.

The models tested in \cite{Pavlovskii2016} were composed of a donor at ZAMS with masses ranging from 0.9 to 1.8$\Msun$, and a NS varying from 1.24 to 1.6$\Msun$. They assumed solar metallicity and Reimer's wind prescription. That study, to find the mass transferred the $L_1$ Lagrange point, $\rm \dot{M}_{\rm d}^{\it L1}$, used the method described in \cite{Pavlovskii2015}, while we use the "Ritter" prescription. The method of \cite{Pavlovskii2015} is important for determining the initial MT stability in systems with a very high mass ratio, and the use of the "Ritter" prescription should not play a role in the test of Sco X-1, or in finding long-lived LMXBs.
\cite{Pavlovskii2016} considered the case of standard MB and wind-boosted MB, adopting $\gamma_{\rm mb}=3$ for both. This is the default value for $\rm \gamma_{mb}$ in {\tt MESA}\footnote{We note this value was chosen to be default by the {\tt MESA} core developers groups, to make a test comparison to results published in the past, and 
is not motivated by physics; we remain convinced that the standard Skumanich law should be used with $\gamma_{\rm mb}=4$.}. \cite{Pavlovskii2016} find the default prescription of MB gives insufficient mass transfer to reproduce the observed mass transfer rate of Sco X-1, by at least an order of magnitude. To produce the observed properties of Sco X-1 (within $\approx 10\%$ estimated uncertainty), a modified MB law must instead be used.

\begin{figure}
    \centering
    \includegraphics[width=\columnwidth]{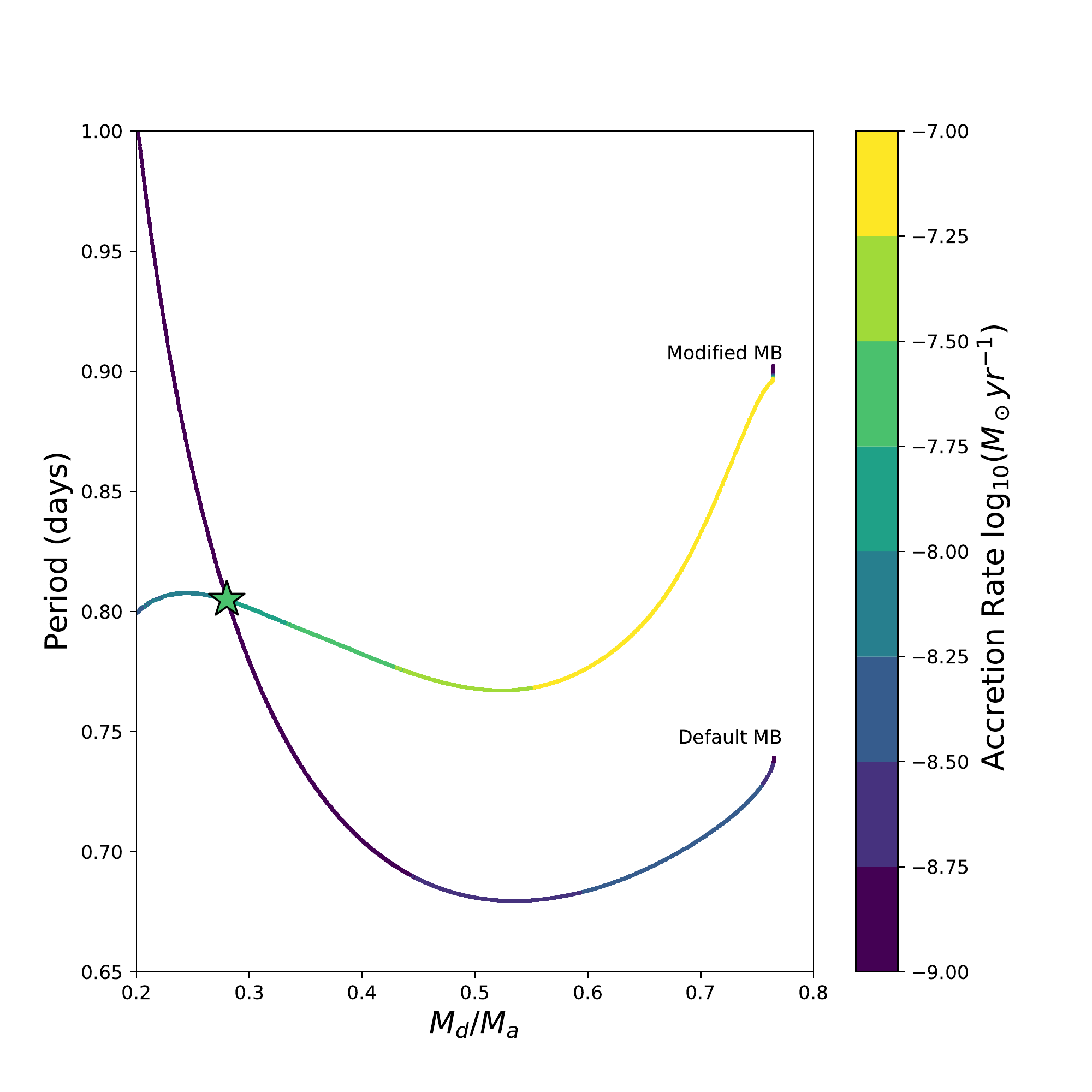}
    \caption{The results of the two models tested to verify our results to reproduce the observed system Sco X-1. The upper curve in the plot is the evolution using the modified MB, and the lower curve is the evolution using the default MB. The star on the plot represents the approximate location of Sco X-1 based on observations with the colour of the star corresponding to the observed mass transfer rate.} The colour bar denotes the MT rate of the binary system with the observed MT of Sco X-1 observed as $\log_{10}(M_\odot \rm / yr^{-1}) \approx -7.7$. The modified MB can reproduce the appropriate MT observed in Sco X-1 while the default MB cannot.
    \label{fig:Sco_Test}
\end{figure}

In this test only, for comparison purposes, we have run similar MT sequences with $\gamma_{\rm mb}=3$. We considered one of the sets of binary companion masses presented in \cite{Pavlovskii2016}: a 1.3$\Msun$ NS and a 1.0 $\Msun$ donor. The test run for the default MB was done using an initial period of 2.7 days, while for the test of the wind-boosted MB we used an initial period of 7.6 days. Both initial periods were taken from systems in \cite{Pavlovskii2016}. The binary systems were evolved until they had similar orbital periods as Sco X-1. The MT tracks of the two systems are shown in Figure \ref{fig:Sco_Test}. They are similar to those shown in Figures 1 and 2 of \cite{Pavlovskii2016}. The results of the simulations with the two MB prescriptions as well as the observed values are listed in Table \ref{table:ScoX1}. 

As can be seen, we 
confirm that 
the modified MB prescription better reproduces the observed value of the MT rate in Sco X-1. We will return to the case of Sco X-1 in section \ref{sec:stat} to review which MB prescriptions can reproduce Sco X-1.

\begin{table}
\caption{\textbf{Sco X-1 Test Properties}}
\centering
\resizebox{0.45\textwidth}{!}{\begin{tabular}{l|cccc}
Quantity & Observed & Boosted MB & Default MB & Ref. \\
\hline
Mass Ratio & 0.28 - 0.51  & 0.42 & 0.28 & (1,2)\\
Period [Hours] & 18.89551  & 18.90 & 18.90 & (3)\\
MT [${M_\odot}{\rm yr}^{-1}$] & $ 2.2 \times 10^{-8} $ & $ 2.1 \times 10^{-8}$ & $ 1.1 \times 10^{-9}$ & (3) \\
Effective Temperature [K] & <4800 & 4718 & 4627 & (1,2) \\
Donor Mass [$M_\odot$] & 0.28 - 0.70 & 0.56 & 0.50 & (1,2)\\
NS Mass [$M_\odot$] & $< 1.73$ & 1.33 & 1.79 & (1,2)\\
\end{tabular}}
\label{table:ScoX1}
\begin{flushleft}
  \textbf{Notes.} The calculated values are taken at the point where the period of the simulated binary is $\approx 18.89551$ hours. (1) \cite{Steeghs2002}; (2) \cite{Sanchez2015}; (3) \cite{Watts2008} 
\end{flushleft}
\end{table}

\section{Observational Data for NS LXMBs}
\label{sec:Observational Data}

\begin{table*} 
\centering
\caption{\textbf{NS Binaries}}
\resizebox{0.95\textwidth}{!}{\begin{tabular}{l|ccccccl}

Source                 & Flags          & $M_d$                                   & Mass Ratio                      & Period                  & Distance                              & Average Mass Transfer                                  & Reference \\
                       &                & ($M_\odot$)                             &  $(M_d/M_a)$                                          & (Hours)                 & (kpc)                                 & ($M_\odot$ yr$^{-1}$)                                  & \\
\hline 
4U 1820-303            & Per, GC, UC    & $ -                                          $ & $ -                                             $ & $ 0.183 \ _{\rm S87}         $ & $ 7.9 \pm 0.4 \ _{\rm H13}                 $ & $ 1.2 \pm 0.6\times 10^{-8} \ _{\rm H13}                    $ & S87, H13 \\
4U 0513-40             & Per, GC, UC    & $ 0.045                                      $ & $ 0.03^{*}                                      $ & $ 0.283 \ _{\rm Z09}         $ & $ 12.1 \pm 0.6 \ _{\rm H13}                $ & $ 1.2 \pm 0.6\times 10^{-9} \ _{\rm H13}                    $ & Z09, H13 \\
2S 0918-549            & Per, UC        & $ 0.024 - 0.039 \ _{\rm Z11}                 $ & $ 0.02 - 0.03^{*}                               $ & $ 0.290 \ _{\rm S01}         $ & $ 5.4 \pm 0.8 \ _{\rm H13}                 $ & $ 2.6 \pm 1.5\times 10^{-10} \ _{\rm H13}                   $ & Z11, H13 \\
4U 1543-624            & Per, UC        & $ 0.03 \ _{\rm W04}                          $ & $ 0.02^{*}                                      $ & $ 0.303 \ _{\rm W04}         $ & $ 7 \ _{\rm H13}                           $ & $ 1.3 \substack{+1.8 \\ -1.2}\times 10^{-9} \ _{\rm H13}    $ & W04, H13 \\
4U 1850-087            & Per, GC, UC    & $ 0.04 \ _{\rm H96}                          $ & $ 0.03^{*}                                      $ & $ 0.343 \ _{\rm H96}         $ & $ 6.9 \pm 0.3 \ _{\rm H13}                 $ & $ 2.2 \pm 1.1\times 10^{-10} \ _{\rm H13}                   $ & H96, H13 \\
M15 X-2                & Per, GC, UC    & $ 0.02-0.03 \ _{\rm D05b}                    $ & $ 0.02^{*}                                      $ & $ 0.377 \ _{\rm D05b}        $ & $ 10.4 \pm 0.5 \ _{\rm H13}                $ & $ 3.8 \pm 1.9\times 10^{-10} \ _{\rm H13}                   $ & D05b, H13 \\
IGR J17062-6143        & T, PSR, UC     & $ 0.0155-0.0175 \ _{\rm S18b}                $ & $ 0.01^{*}                                      $ & $ 0.633 \ _{\rm S18b}        $ & $ 7.3 \pm 0.5 \ _{\rm K17}                 $ & $ 2.5 \times10^{-11} \ _{\rm K17}                           $ & K17, S18b \\
XTE J1807-294          & T, PSR, UC     & $ -                                          $ & $ -                                             $ & $ 0.670 \ _{\rm M03}         $ & $ 8 \substack{+4 \\ -3.3} \ _{\rm H13}     $ & $ <1.5 \substack{+1.9 \\ -1.2}\times 10^{-11} \ _{\rm H13}  $ & M03, H13 \\
4U 1626-67             & Per, PSR, UC   & $ <0.036 \ _{\rm H13}                        $ & $ 0.02^{*}                                      $ & $ 0.700 \ _{\rm C98a}        $ & $ 8 \substack{+5 \\ -3} \ _{\rm H13}       $ & $ 8.0 \substack{+14 \\ -6}\times 10^{-10} \ _{\rm H13}      $ & C98a, H13 \\
XTE J1751-305          & T, PSR, UC     & $ -                                          $ & $ -                                             $ & $ 0.710 \ _{\rm M02}         $ & $ 8 \substack{+0.5 \\ -1.3} \ _{\rm H13}   $ & $ 5.1 \substack{+2.6 \\ -2.9}\times 10^{-12} \ _{\rm H13}   $ & M02, H13 \\
XTE J0929-314          & T, PSR, UC     & $ \simeq 0.01 \ _{\rm G02}                   $ & $ 0.007                                         $ & $ 0.730 \ _{\rm G02}         $ & $ 8 \substack{+7 \\ -3} \ _{\rm H13}       $ & $ <9.7 \substack{+25 \\ -7.7}\times 10^{-12} \ _{\rm H13}   $ & G02, H13 \\
4U 1916-053            & Per, UC        & $ 0.064\pm 0.01 \ _{\rm H13}                 $ & $ 0.046                                         $ & $ 0.833 \ _{\rm W81}         $ & $ 9.3 \pm 1.4 \ _{\rm H13}                 $ & $ 6.3 \pm 3.7\times 10^{-10} \ _{\rm H13}                   $ & W81, H13 \\
Swift J1756.9-2508     & T, PSR, UC     & $ \leq 0.022 \ _{\rm K07}                    $ & $ 0.02                                          $ & $ 0.912 \ _{\rm K07}         $ & $ 8 \pm 4 \ _{\rm H13}                     $ & $ 1.7 \substack{+2.3 \\ -1.5} \times 10^{-11} \ _{\rm H13}  $ & K07, H13 \\
NGC 6440 X-2           & T, PSR, GC, UC & $ \simeq 0.0076 \ _{\rm B15}                 $ & $ 0.005                                         $ & $ 0.955 \ _{\rm A10}         $ & $ 8.5 \pm 0.4 \ _{\rm H13}                 $ & $ 1.0 \pm 0.5\times 10^{-12} \ _{\rm H13}                   $ & A10, B15, H13 \\
HETE J1900.1-2455      & T, PSR         & $ 0.016 - 0.07 \ _{\rm K06a}                 $ & $ 0.01 - 0.05^{*}                               $ & $ 1.39 \ _{\rm W08}          $ & $ 4.7 \pm 0.6 \ _{\rm W08}                 $ & $ 4.63 \times 10^{-11} \ _{\rm W08}                         $ & K06a, W08 \\
1A 1744-361            & T              & $ 0.07-0.22 \ _{\rm B06}                     $ & $ 0.07?^{*}                                     $ & $ 1.62 \pm 0.37 \ _{\rm W08} $ & $ 6 \pm 3 \ _{\rm W08}                     $ & $ 2.22 \times 10^{-11} \ _{\rm W08}                         $ & B06, W08 \\
IGR J17379-3747        & T, PSR         & $ >0.056 \ _{\rm S18a}                       $ & $ > 0.04^{*}                                    $ & $ 1.88 \ _{\rm S18a}         $ & $ \sim 8.5 \ _{\rm S18a}                   $ & $ \sim 4 \times 10^{-11} \ _{\rm S18a}                      $ & S18a, S18c \\ 
SAX J1808-3658         & T, PSR         & $ 0.04\substack{+0.02 \\ -0.01} \ _{\rm W13} $ & $ 0.04^{*}                                      $ & $ 2.01 \ _{\rm C98b}         $ & $ 3.4-3.6 \ _{\rm C12}                     $ & $ 1.73 \times 10^{-11} \ _{\rm C12}                         $ & C98b, C12, W13 \\
XB 1832-330            & T,GC           & $ -                                          $ & $ -                                             $ & $ 2.1 \ _{\rm E12}           $ & $ 10.0 \ _{\rm P01}                        $ & $ \sim 3 \times10^{-10} \ _{\rm E12}                        $ & P01, E12 \\
IGR 00291+5934         & T, PSR         & $ 0.039-0.16 \ _{\rm D17}                    $ & $ 0.02-0.11                                     $ & $ 2.46 \ _{\rm G06}          $ & $ 2.6-3.6 \ _{\rm C12}                     $ & $ 1 \times 10^{-12} \ _{\rm D17}                            $ & C12, G05, D17 \\
4U 1822-00             & Per, M         & $ -                                          $ & $ -                                             $ & $ 3.20? \ _{\rm S07}         $ & $ 6.3 \pm 2 \ _{\rm S07}                   $ & $ 9.0 \substack{+8.0 \\ -5.0} \times 10^{-10} \ _{\rm S07}  $ & S07 \\
4U 1636-536            & Per            & $ 0.29-0.48^{*}                              $ & $ 0.21-0.34 \ _{\rm W16}                        $ & $ 3.79 \ _{\rm W08}          $ & $ 6 \pm 0.5 \ _{\rm W08}                   $ & $ 1.25 \times 10^{-9} \ _{\rm C12}                          $ & C12, W08, W16 \\
EXO 0748-676           & T              & $ 0.1? \ _{\rm D14a}                         $ & $ 0.07?^{*}                                     $ & $ 3.82 \ _{\rm D14a}         $ & $ 7.1 \pm 1.2 \ _{\rm D14a}                $ & $ <4.4 \times 10^{-10} \ _{\rm C12}                         $ & C12, D14a \\
4U 1254-69             & Per            & $ 0.45 \ _{\rm C68b}                         $ & $ 0.33-0.36 \ _{\rm C13}                        $ & $ 3.93 \ _{\rm W08}          $ & $ 13 \pm 3 \ _{\rm W08}                    $ & $ 1.77 \times 10^{-9} \ _{\rm W08}                          $ & C86b, C13, W08 \\
4U 1728-16 (GX 9+9)    & Per            & $ 0.4 \ _{\rm K09b}                          $ & $ 0.29 \ _{\rm K09b}                            $ & $ 4.20 \ _{\rm L07}          $ & $ 5? \ _{\rm K06b}                         $ & $ 2.91 \times 10^{-9} \ _{\rm C97}                          $ & C97, K06b, L07 \\
XTE J1814-338          & T              & $ 0.19-0.32 \ _{\rm W17}                     $ & $ 0.123 \substack{+0.012 \\ -0.01} \ _{\rm W17} $ & $ 4.27 \ _{\rm W08}          $ & $ 8\pm 1.6 \ _{\rm C12}                    $ & $ <5.99 \times 10^{-12} \ _{\rm C12}                        $ & C12, W08, W17 \\
4U 1735-444            & Per            & $ \leq 0.58                                  $ & $ 0.05-0.41^{C06}                               $ & $ 4.65 \ _{\rm W08}          $ & $ 8.5 \pm 1.3 \ _{\rm W08}                 $ & $ 6.31 \times 10^{-9} \ _{\rm C12}                          $ & C06, C12, W08 \\
4U 1746-37             & Per, GC        & $ --                                         $ & $ -                                             $ & $ 5.16 \ _{\rm B04}          $ & $ 11.6 \ _{\rm B04}                        $ & $ 1 \times10^{-9} \ _{\rm B04}                              $ & S01, B04\\
2A 1822-371            & Per, M         & $ 0.47 \pm 0.04 \ _{\rm I15}                 $ & $ 0.28 \ _{\rm I15}                             $ & $ 5.57 \ _{\rm B17}          $ & $ 2.5 \ _{\rm B17}                         $ & $ \sim 2 \times 10^{-8} \ _{\rm B17}                        $ & B17, I15 \\
XTE J2123-058          & T              & $ 0.76 \pm 0.22 \ _{\rm C02}                 $ & $ 0.49 \pm 0.1 \ _{\rm C01}                     $ & $ 5.96 \ _{\rm W08}          $ & $ 9.6 \pm 1.3 \ _{\rm W08}                 $ & $ <7 \times 10^{-12} \ _{\rm C12}                           $ & C02, C12, S03a, W08 \\
X 1658-298             & T              & $ 0.3 - 0.8 \ _{\rm P18}                     $ & $ 0.21 - 0.57^{*}                               $ & $ 7.12 \ _{\rm D14a}         $ & $ 12 \pm 3 \ _{\rm W08}                    $ & $ 1 \times 10^{-9} \ _{\rm W08}                             $ & D14a, P18, W08 \\
2A 0521-720 (LMC X-2)  & Per, M         & $ -                                          $ & $ -                                             $ & $ 8.16 \ _{\rm L07}          $ & $ 50 \pm 2 \ _{\rm A09}                    $ & $ 4 \times 10^{-8} \ _{\rm C12}                             $ & C12, A09, L07 \\
SAX J1748.9-2021       & T, GC          & $ 0.12 - 1 \ _{\rm S16}                      $ & $ 0.09 - 0.71^{*}                               $ & $ 8.76 \ _{\rm S16}          $ & $ 8.5 \pm 0.4 \ _{\rm S16}                 $ & $ \sim 7 \times 10^{-11} \ _{\rm W08}                       $ & S16, W08 \\
IGR J18245-2452        & T, GC          & $ > 0.17 \ _{\rm P13}                        $ & $ 0.12^{*}                                      $ & $ 11.0 \ _{\rm P13}          $ & $ 5.5 \ _{\rm P13}                         $ & $ \lesssim 1.0 \times 10^{-10} \ _{\rm P13}                 $ & P13 \\ 
GRS 1747-312           & T, GC          & $ --                                         $ & $ -                                             $ & $ 12.36 \ _{\rm I03}         $ & $ 9.5 \ _{\rm V18}                         $ & $ 1 \times10^{-10} \ _{\rm V18}                             $ & B04, V18 \\
4U 1456-32 (Cen X-4)   & T              & $ 0.31 \pm 0.27 \ _{\rm D05a}                $ & $ 0.18 \pm 0.06 \ _{\rm D05a}                   $ & $ 15.1 \ _{\rm L07}          $ & $ 1.2 \pm 0.2 \ _{\rm C12}                 $ & $ 4 \times 10^{-11} \ _{\rm C12}                            $ & C12, D05a, L07 \\
AC 211 (X2127+12)      & Per, GC        & $ \sim 0.1 \ _{\rm V04}                      $ & $ \sim 0.1 \ _{\rm V04}                         $ & $ 17.1 \ _{\rm I93}          $ & $ 10.4 \ _{\rm C68b}                       $ & $ \sim 7 \times10^{-9} \ _{\rm I93}                         $ & C86, I93, V04 \\ 
H 1617-155 (Sco X-1)   & Per, M         & $ 0.28 - 0.70 \ _{\rm S15}                   $ & $ 0.28 - 0.51 \ _{\rm S15}                      $ & $ 18.9 \ _{\rm W08}          $ & $ 2.8 \pm 0.3 \ _{\rm S15}                 $ & $ 3 \times 10^{-8} \ _{\rm C12}                             $ & C12, S15, W08 \\
4U 1908+005 (Aql X-1)  & T              & $ -                                          $ & $ -                                             $ & $ 18.9 \ _{\rm W08}          $ & $ 4.55 \pm 1.35 \ _{\rm W08}               $ & $ 6 \times 10^{-10} \ _{\rm C12}                            $ & C12, W08 \\
4U 1624-49             & Per            & $ -                                          $ & $ -                                             $ & $ 20.9 \ _{\rm B00}          $ & $ 15 \substack{+2.9 \\ -2.6} \ _{\rm X09}  $ & $ 4.6 \times 10^{-9} \ _{\rm B09}                           $ & B09, L05, X09 \\ 
3A 1702-363 (GX 349+2) & Per            & $ 0.78^{*}                                   $ & $ \sim 0.56 \ _{\rm I09}                        $ & $ 21.9 \pm 0.4 \ _{\rm I09}  $ & $ 5 \pm 1.5 \ _{\rm C12}                   $ & $ 2.37 \pm 1 \times 10^{-8} \ _{\rm C12}                    $ & C12, I09, W08 \\
2A 1655+353 (Her X-1)  & T              & $ 2.03 \pm 0.37                              $ & $ 1.45                                          $ & $ 40.8 \ _{\rm L07}          $ & $ 6.1 \substack{+0.9 \\ -0.4} \ _{\rm L14} $ & $ 1.3 \times 10^{-8} \ _{\rm C12}                           $ & C12, L07, R11 \\
4U 2142+38 (Cyg X-2)   & Per, M         & $ 0.56 \pm 0.07 \ _{\rm P16}                 $ & $ 0.34 \pm 0.01 \ _{\rm P16}                    $ & $ 236.3 \ _{\rm W08}         $ & $ 10.55 \pm 4.45 \ _{\rm W08}              $ & $ 3.0 \times 10^{-8} \ _{\rm C12}                           $ & C12, M18, W08 \\
GRO J1744-28           & T, M           & $ 0.2 - 0.7 \ _{\rm D15}                     $ & $ 0.15 - 0.5^{*}                                $ & $ 284.0 \ _{\rm L07}         $ & $ 8? \ _{\rm D15}                          $ & $ \sim 1\times 10^{-8} \ _{\rm D14b}                        $ & D15, D14b, L07 \\

\end{tabular}}
\label{NS_Table}
\begin{flushleft}
\textbf{Notes.} The periods and mass transfer rates are either taken directly from the listed reference, or calculated using values from that reference. Any value with a ? attached is a rough estimate of that value. If possible, the mass fraction or companion mass is taken directly from the source. If a  companion mass or mass fraction can be calculated, this is done assuming a neutron star mass of 1.4M$_\odot$. These calculated values are denoted by a $^{*}$ and are simply an approximation. We include any errors that could be taken directly from the reference paper. In cases where an error was not listed but can be calculated, we did so. The error in the mass transfer rate is calculated by looking at the errors in distances, flux, or luminosity from the listed reference. The second column notes systems that are persistent (Per), transient (T),  with a neutron star of mass exceeding $1.6M_\odot$ (M), systems with pulsars (PSR), systems in globular clusters (GC), and ultra compact systems (UC).
References: 
A09 - \cite{Agrawal2009}, A10 - \cite{Altamirano2010}, B00 - \cite{BalucinskaChurch2004}, B04 - \cite{BalucinskaChurch2004}, B06 - \cite{Bhattacharyya2006}, B09 - \cite{Balman2009}, B10 - \cite{Bayless2010}, B15 - \cite{Bult2015}, B17 - \cite{Bak2017}, C86a - \cite{Charles1986} C86b - \cite{Courvoisier1986}, C97 - \cite{Christian1997}, C98a - \cite{Chakrabarty1998a}, C98b - \cite{Chakrabarty1998b}, C02 - \cite{Casares2002}, C06 - \cite{Casares2006}, C12 - \cite{Coriat2012}, C13 - \cite{Cornelisse2013}, D05a - \cite{Davanzo2005}, D05b - \cite{Dieball2005}, D14a - \cite{DAi2014}, D14b - \cite{Degenaar2014}, D15 - \cite{DAi2015}, D17 - \cite{deFalco2017}, E12 - \cite{Engel2012}, G02 - \cite{Galloway2002}, G05 - \cite{Galloway2005}, H96 - \cite{Homer1996}, H07 - \cite{Heinke2007}, H10 - \cite{Harris2010}, H13 - \cite{Heinke2013}, I93 - \cite{Ilovaisky1993}, I03 - \cite{Int2003}, I09 - \cite{Iaria2009}, I15 - \cite{Iaria2015}, J10 - \cite{Jain2010}, K06a - \cite{Kaaret2006}, K06b - \cite{Kong2006}, K07 - \cite{Krimm2007}, K17 - \cite{Keek2017}, L05 - \cite{Lommen2005}, L07 - \cite{Liu2007}, M02 - \cite{Markwardt2002}, M03 - \cite{Markwardt2003}, M18 - \cite{Mondal2018}, P01 - \cite{Parmar2001}, P13 - \cite{Papitto2013}, P16 - \cite{Premachandra2016}, P17 - \cite{Patruno2017}, P18 - \cite{Ponti2018}, R11 - \cite{Rawls2011}, S01 - \cite{Sidoli2001}, S87 - \cite{Stella1987}, S03a - \cite{Shahbaz2003}, S07 - \cite{Shahbaz2007}, S15 - \cite{Sanchez2015} S16 - \cite{Sanna2016}, S18a - \cite{Sanna2018}, S18b - \cite{Strohmayer2018a}, S18c - \cite{Strohmayer2018b} W81 - \cite{Walter1981}, W04 - \cite{Wang2004}, W08 - \cite{Watts2008}, W13 - \cite{Wang2013}, W16 - \cite{Wisniewicz2016}, W17 - \cite{Wang2017}, V04 - \cite{VanZyl2004}, V18 - \cite{Vats2018}, X09 - \cite{Xiang2009}, Z09 - \cite{Zurek2009}, Z11 - \cite{Zhong2011} 
\end{flushleft}
\end{table*}

To compare our results to observations, we have compiled a list of up-to-date properties of some Milky Way NS LMXBs. See Table~\ref{NS_Table} for a  list of systems, relevant data, and references.

The systems with periods shorter than 80 minutes are ultra-compact X-ray binaries (UCXBs), where the donor star must be partly or completely degenerate.
Our binary evolution method may not necessarily be the dominant method to produce these systems, but we keep them in the consideration. Within the table there are also systems labelled as GC systems, which are systems found in globular clusters. Binaries formed within a globular cluster are not likely to be produced from primordial binaries, but instead are more effectively produced via dynamical encounters between binaries, as well as via physical collisions between NSs and subgiants \citep{Verbunt1987, Ivanova2005, Ivanova2008}.

The other flags shown in the second column denote if the source is a persistent or transient system. Persistent systems are those where bright ($L_X>10^{35}$ erg/s) X-ray emissions have been consistently seen whenever X-ray monitoring missions have observed these systems, over a 40-60 year timespan. Transient systems have large changes in their X-ray emission, typically exceeding $L_X=10^{36}$ erg/s at some points (outbursts) and declining below $L_X=10^{35}$ erg/s at other times (quiescence). There are several systems currently thought to be persistent which could instead be in a long-term outburst state (lasting $>$50 years), and may be reclassified as transient systems in the future.

\cite{Meyer1981} predicted the existence of a critical MT rate separating the persistent and transient systems. The disc instability model (DIM) predicts 
under what circumstances an accretion disk will experience instabilities \citep[see][for a review of DIM]{Lasota2001}. Stability in the context of the DIM means an accretion disc does not experience outbursts; a stable disc that remains hot will produce a persistent X-ray binary. The criterion for stability is given by \citet{Coriat2012}:

\begin{equation}
\dot{M}_{\rm{crit}} = kP_{\rm hr}^b \ \rm g\ s^{-1}
\end{equation}

\noindent Here, $\rm P_{\rm hr}$ is the period of the system in hours. For the non-irradiated case, $k = (2.6 \pm 0.9) \times 10^{16}$, and $b = 1.76$.  If there is irradiation of the accretion disc, and the accretor is a NS, $b = 1.59$ and $k = (2.9 \pm 0.9) \times 10^{15}$. Systems that lie above the DIM line are expected to be persistent while systems below the line are transient.

Here, we only consider the systems where the MT rate is known. If the cited papers do not provide the MT rate but instead provide the X-ray luminosity, $L_x$, we find $\dot{M}$ using

\begin{equation}
\dot{M} = \frac{L_{\rm X}R_{\rm a}}{GM_{\rm a}} \ .
\end{equation}

\noindent
Observationally, an upper limit for NS mass has been found to be $2.01 \pm 0.04 M_\odot$ \citep{Antoniadis2013}, our calculations will be done assuming the mass of the NS is $M_{\rm a}=1.4~M_\odot$ with a radius of $R_{\rm a}=11.5$ km \citep{Ozel2016}. The key properties of interest for this work are the mass ratio $q = M_d/M_a$, the period $p$ and the average mass transfer rate $\dot{M}$.


\section{Principal Results}
\label{sec:Principle Results}


\subsection{Evolutionary Tracks}

\begin{figure*}
    \centering
    \includegraphics[width=\textwidth]{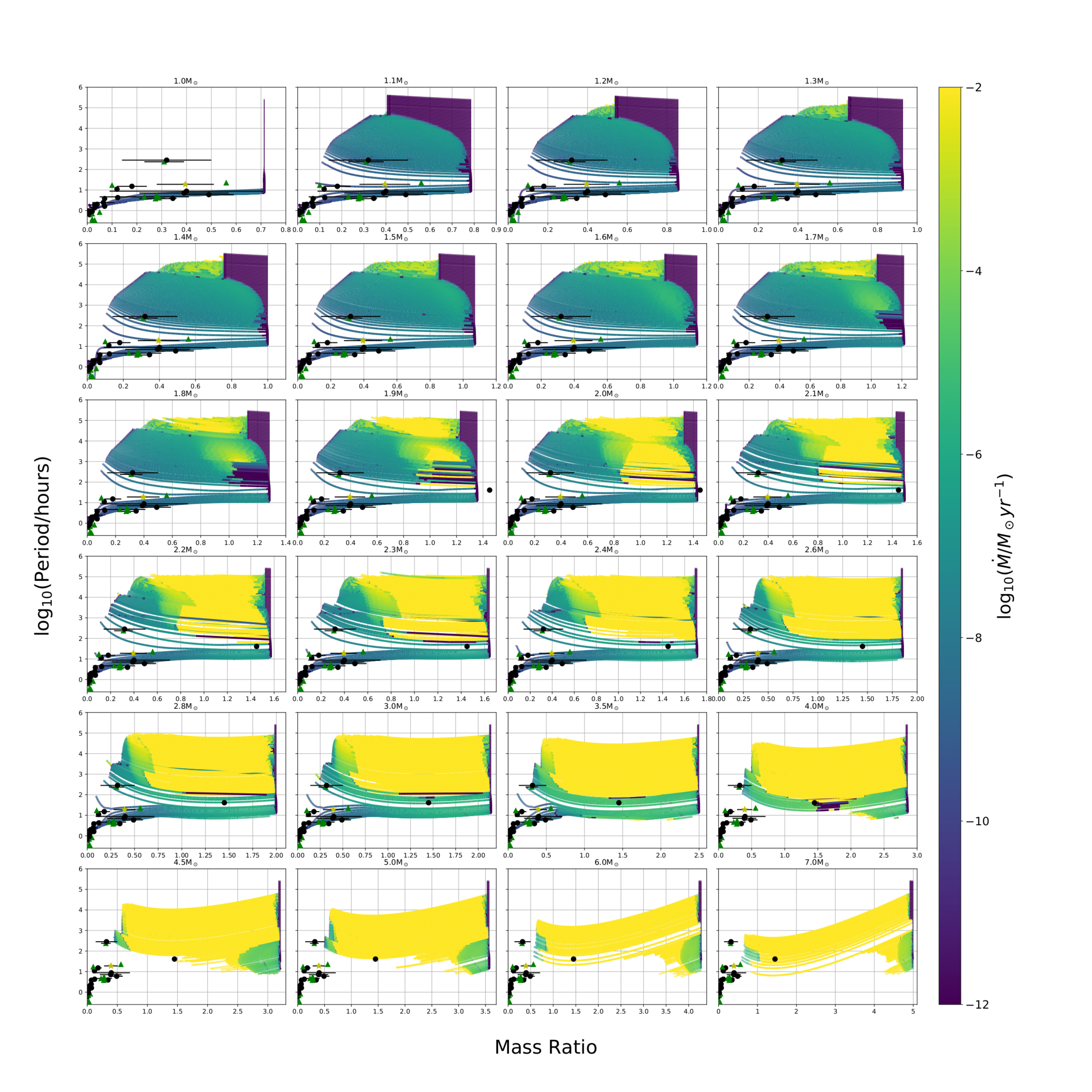}
    \caption{A collection of subplots showing the evolution through mass transfer for all initial periods, shown on the grid of initial donor masses and mass ratios. The data shown is only after the onset of RLOF with the colour bar indicating log$_{10}$(MT). The evolutionary tracks evolve from the right of the subplots leftwards as the donor loses mass through mass transfer. At higher donor masses and low periods, there exist a subset of systems which abruptly terminate their tracks as they begin to transfer mass dynamically. The points on the plot represent observable systems, with errors found in Table \ref{NS_Table}. Circles are transient systems, and triangles are persistent systems. The single star point is the binary Sco X-1 used to test the validity of our numeric results. The range of periods in each plot is the same, but the range of donor masses differs. The abrupt cutoff at higher periods is a result of the star reaching the end of its life.}
    \label{fig:NB_mass_grid}
\end{figure*}

\begin{figure*}
    \centering
    \includegraphics[width=\textwidth]{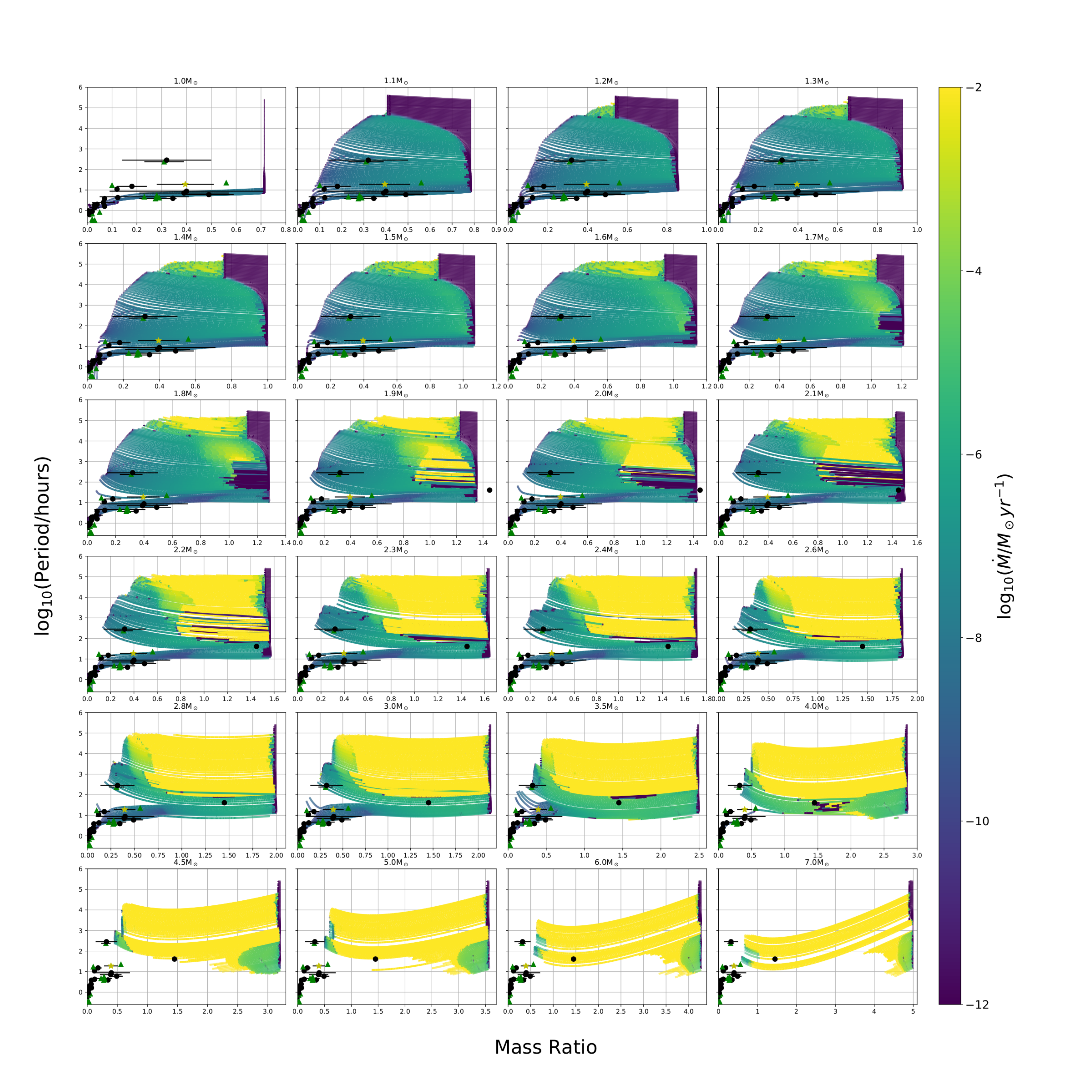}
    \caption{A similar figure to \ref{fig:NB_mass_grid} using case 2, convection boosted MB from Table \ref{tab:MB_scalings}.
    The general behaviour of these simulated systems is similar to that of the systems following the Skumanich law, in Fig. \ref{fig:NB_mass_grid}.}
    \label{fig:CB_mass_grid}
\end{figure*}

\begin{figure*}
    \centering
    \includegraphics[width=\textwidth]{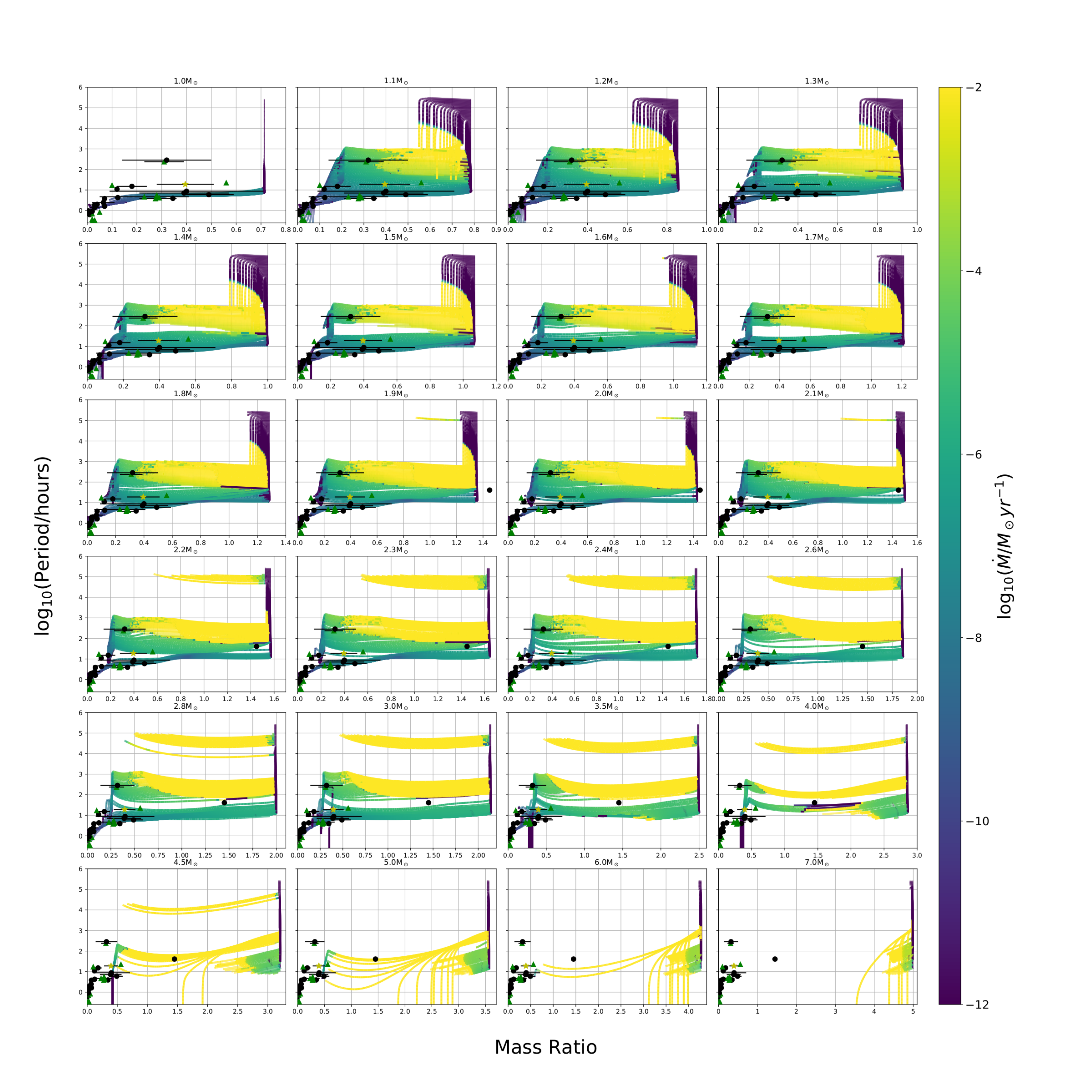}
    \caption{A similar figure to \ref{fig:NB_mass_grid} using case 3, intermediate MB from Table \ref{tab:MB_scalings}. Note that the mass transfer rates are substantially higher than in the previous cases, Figs. \ref{fig:NB_mass_grid} and \ref{fig:CB_mass_grid}.}
    \label{fig:Int_mass_grid}
\end{figure*}

\begin{figure*}
    \centering
    \includegraphics[width=\textwidth]{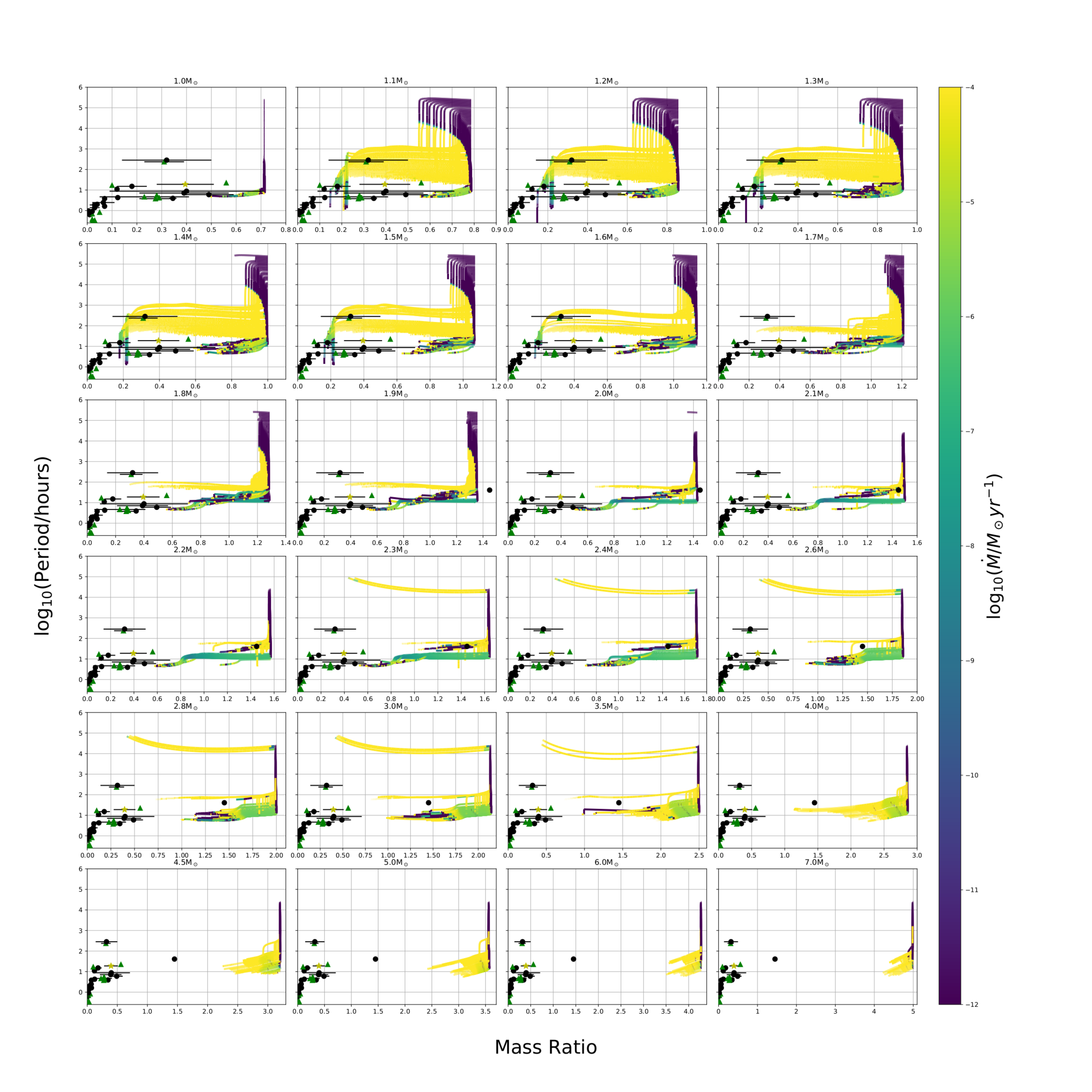}
    \caption{A similar figure to \ref{fig:NB_mass_grid} using case 4, wind boost MB from Table \ref{tab:MB_scalings}. Here the mass transfer rates are extremely high.}
    \label{fig:WB_mass_grid}
\end{figure*}

Here we present the results for the 2136 binary models of each MB prescription by plotting the evolutionary tracks grouped by donor mass on subplots in figures \ref{fig:NB_mass_grid} - \ref{fig:WB_mass_grid}. These figures show the donor mass and period evolution of each simulated binary over the course of its lifetime. Mass in these figures is meant as a proxy for time as the donor stars all decrease in mass over the course of the binary evolution. The binary population simulations each required on order of hours to days to finish, totalling approximately thirty core years of simulation time. For comparison purposes, we overlay the observed data points from Table \ref{NS_Table} on the plots. Sco X-1 is denoted with a star in the subplots as it was our test case from \ref{subsec:ScoX1}.

We can see that our choice of initial conditions affects the resulting evolutionary track of the system. In general, the more massive the donor star, the higher the MT rate. The choice of initial period results in changes that are less monotonic in comparison. Many systems starting with short periods may initiate Roche lobe overflow (RLOF) immediately, while longer period systems may lose significant mass through winds before this.

The separation between systems which evolve to longer orbital periods during their evolution, and those in which the orbital periods shrink, is known as the bifurcation period \citep{Podsiadlowski2002, VanderSluys2004}. As the initial donor mass increases, the bifurcation period also increases until we encounter systems undergoing dynamically unstable MT. We will refer to a binary with the orbital period near the end of its evolution larger than the period at the start of RLOF as growing in period, while systems with a shorter period near the end of their evolution are shrinking. One can see that most of the observed LMXBs are located in the region of shrinking systems. Specifically, in the case using the default Skumanich law, several of them are in the parameter space where the shrinking and growing systems bifurcate.

\subsubsection{Default}
The default Skumanich MB results are shown in Figure \ref{fig:NB_mass_grid}. This MB prescription produces a clear bifurcation in periods through the low and intermediate mass donors. The bifurcation period at RLOF is $\sim 1$ day and shows a distinct split in periods as no binaries pass through the region $q < 0.4$, $ 1 \lesssim \log_{10}(P) \lesssim 2.5 $. The sharp transition seen in the low mass high period systems is a result of the binary reaching a stopping conditions given in Section \ref{sec:Binary_calculations}.

\subsubsection{Convection Boosted}
From Figure \ref{fig:CB_mass_grid}, we cannot see many differences between the default MB and the ``convection boosted" MB. One difference between the default and convection boosted MB schemes is that  evolutionary tracks run through the region $q < 0.4$, $ 1 \lesssim \log_{10}(P) \lesssim 2.5 $. The additional systems passing through this region cause some ambiguity in determining bifurcation periods in systems with initial donor masses $M \lesssim 1.5 M_\odot$, since the binaries near the bifurcation period show very little change in period over their evolution.

\subsubsection{Intermediate}
The ``Intermediate" case includes the additional scaling factor which accounts for the effects of  wind mass loss as seen in equation \ref{eq:mod_MB}. Figure \ref{fig:Int_mass_grid} shows the behaviour of the binary systems with this MB prescription. The additional wind scaling plays a significant role in wider binary systems. The stronger MB scheme brings the binary systems together on a shorter timescale. In these systems, the MB and total angular momentum loss are consistently an order of magnitude higher than the default case. Therefore,  the system loses enough angular momentum due to MB that gravitational radiation comes into play once MB stops. 

\subsubsection{Wind Boosted}
\begin{figure}
    \centering
    \includegraphics[width=\columnwidth]{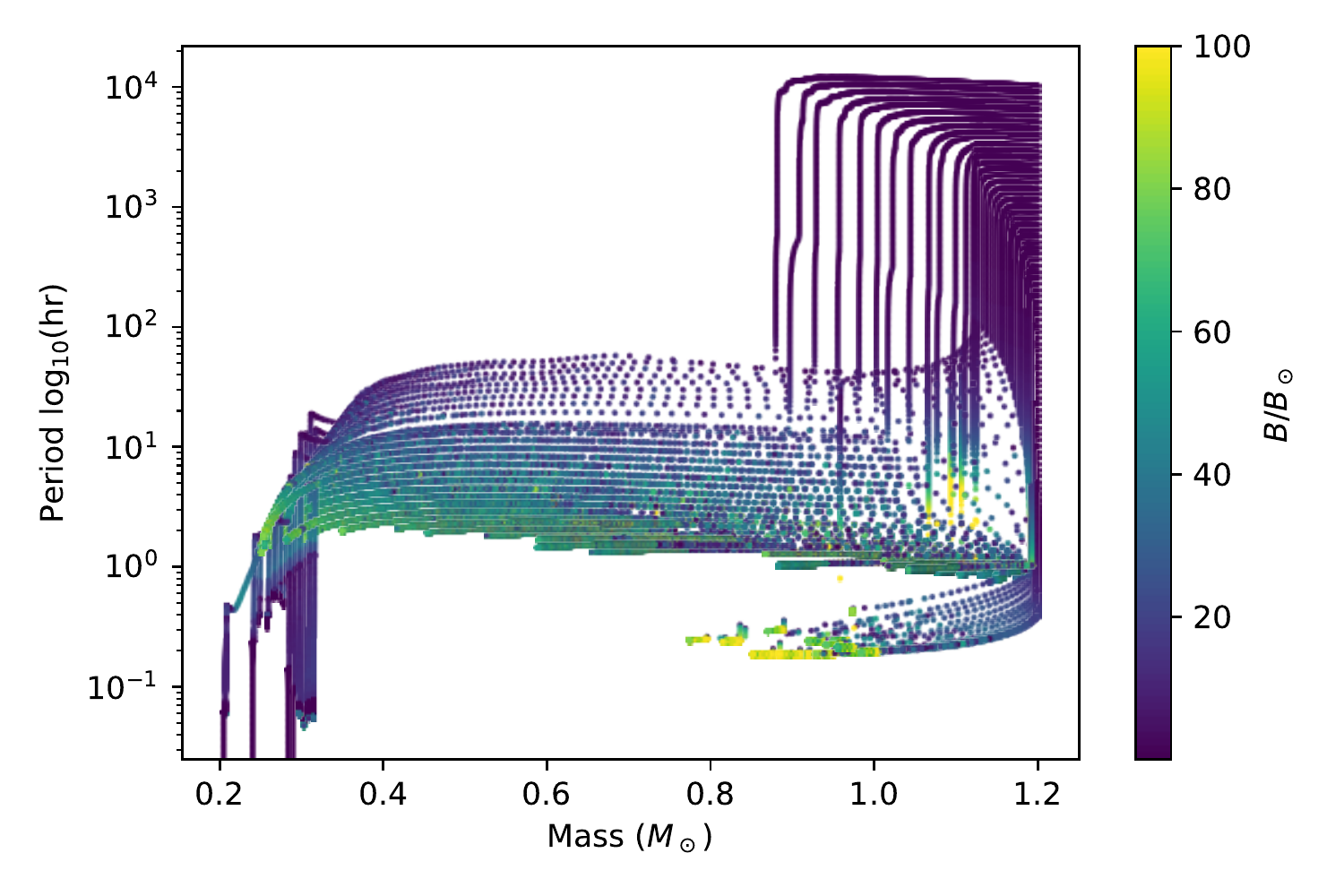}
    \caption{A figure showing the evolutionary tracks of a binary system consisting of a $1.2M_\odot$ initial mass donor with a $1.4M_\odot$ NS at a variety of initial periods using the wind boosted MB. The colour bar shows the magnetic field strength of the donor star at a given point in the binary evolution. The magnetic field is given in units of the solar magnetic field, which is $\sim 1 \rm G$.}
    \label{fig:B_field}
\end{figure}

The ``wind boost" case shown in figure \ref{fig:WB_mass_grid}, includes the effects of the convective turnover time and accounts for the rotation rate of the star. In this case, the individual values of magnetic field strength as calculated using equation \ref{eq:B_scaling}, the turnover time, and the wind mass loss rates are all within reasonable ranges. The magnetic field, which reaches a maximum of 100G shown in Figure \ref{fig:B_field}, is within the range expected for giant stars \citep{Auriere2015}. The convective turnover time is also similar to those calculated by \cite{Pavlovskii2015} for systems that are predicted to reproduce Sco X-1. It appears that the individual properties are all within reasonable ranges. However, the combination of all these boost factors produces MB that is too strong, resulting in MT that consistently exceeds $1M_\odot \ \rm yr^{-1}$. MT at these rates results in evolution on a dynamical timescale, and as such the results from this highly boosted MB should not be trusted. It is likely that this prescription has reached and exceeded a saturation limit that is not accounted for in this work \citep{Mestel1987}. It has been shown that in cases where the rotation rate is high, additional magnetic field structure effects must be included to dampen the angular momentum loss \citep{Ivanova2003}. For completeness, we include the results from the wind-boosted case, but since the MT rate is so high, it is unlikely these simulations 
accurately describe reality.

\subsection{Binary Properties}
\label{subsec:Chap5_Bin_Props}

\begin{figure*}
    \centering
    \includegraphics[width=0.45\textwidth]{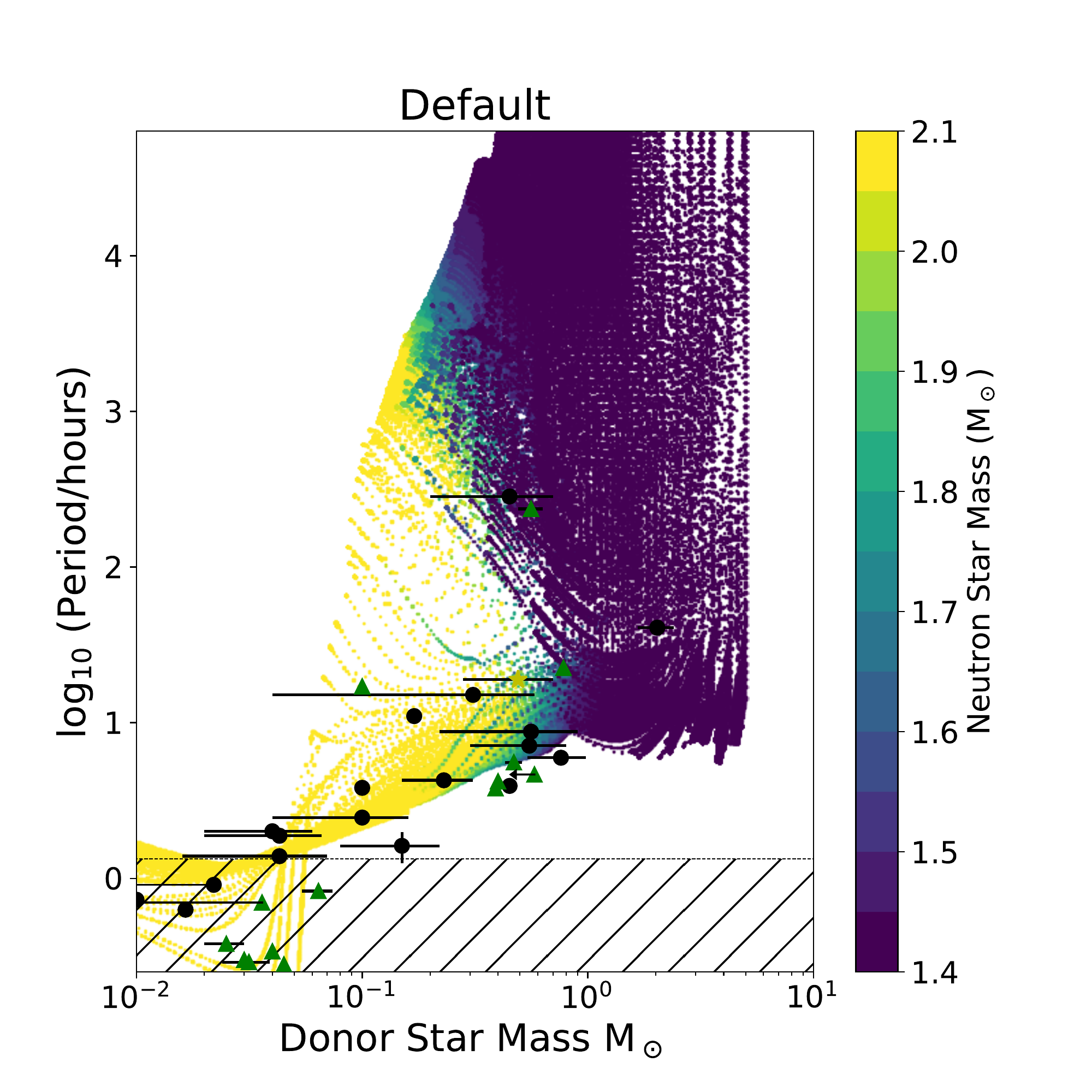}
    \hspace*{0.01cm}
    \includegraphics[width=0.45\textwidth]{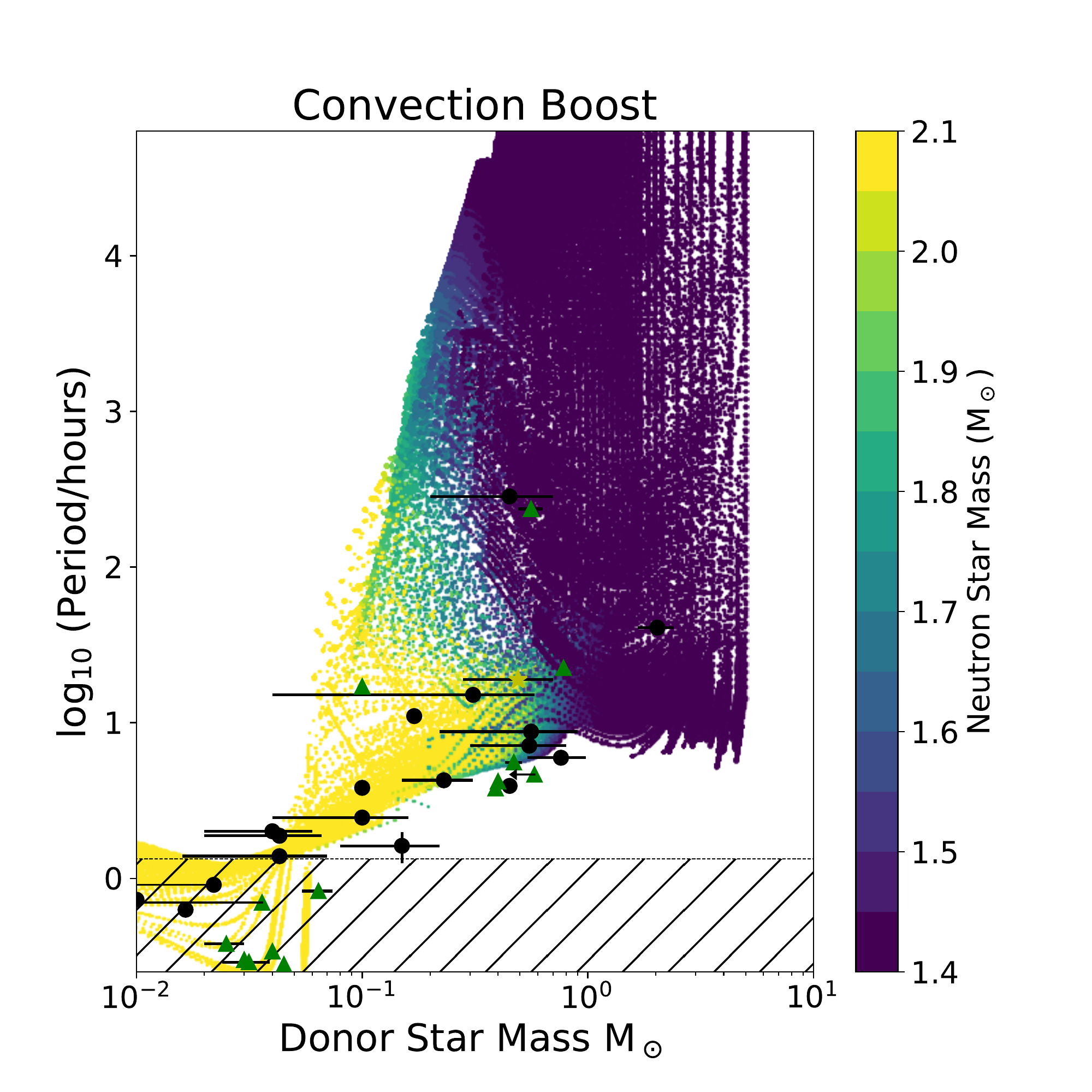}
    \includegraphics[width=0.45\textwidth]{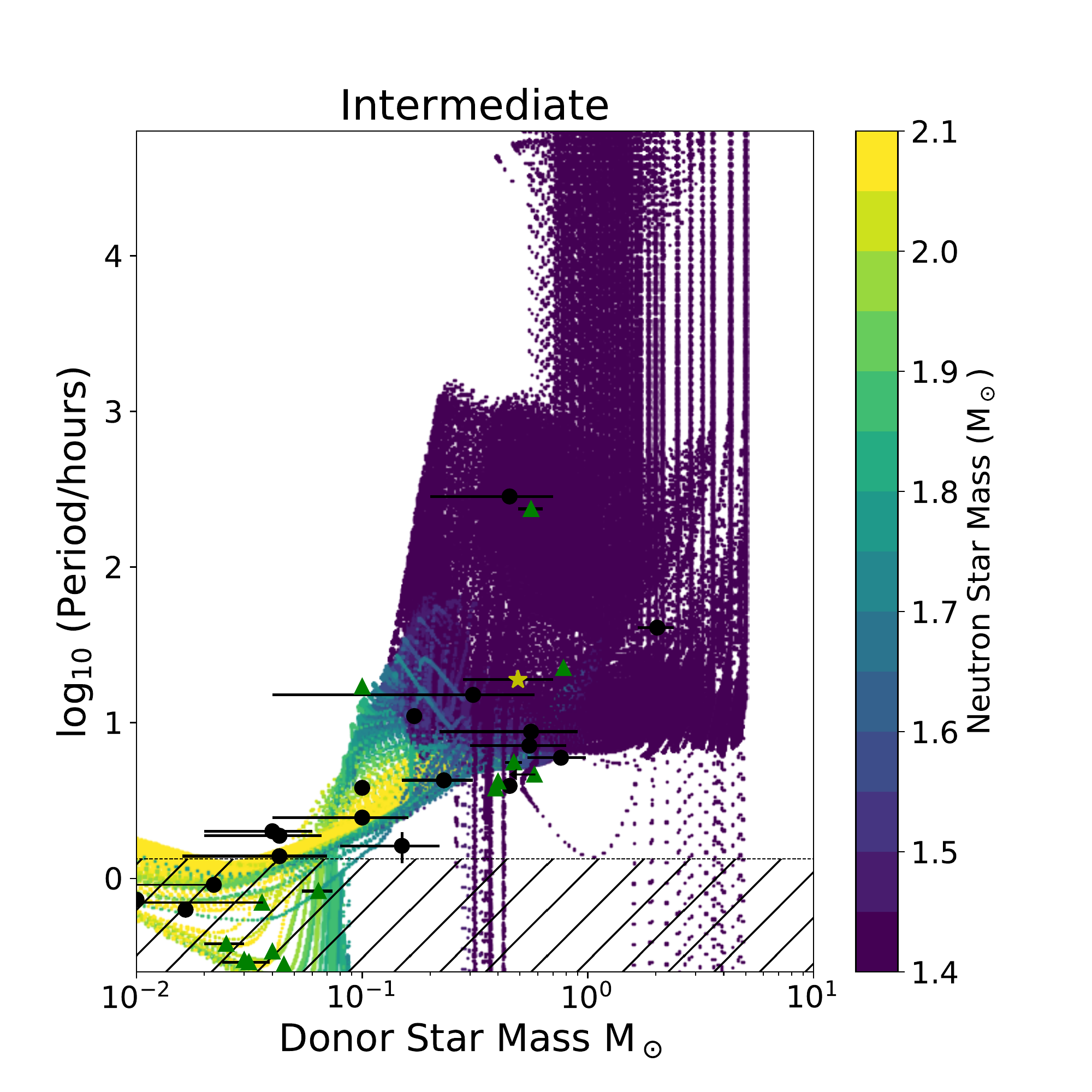}
    \hspace*{0.01cm}
    \includegraphics[width=0.45\textwidth]{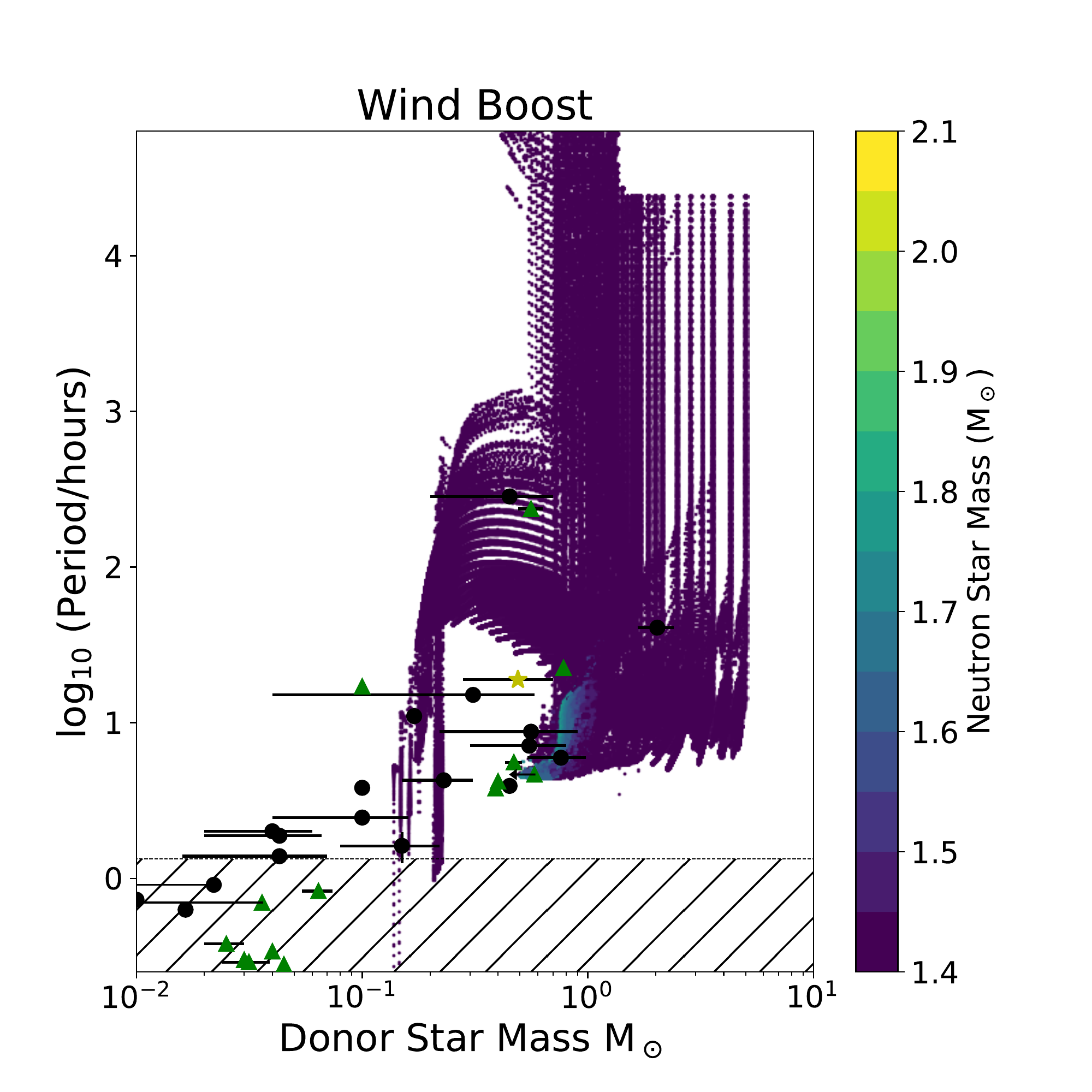}
    \caption{A scatter plot showing the changes in NS mass during the evolution of our systems. The lower hashed area shows the periods corresponding to ultra compact sources. The default and convection boosted cases suggest that all short period binaries contain a massive NS.}
    \label{fig:Mass_Scatter}
\end{figure*}

Figure~\ref{fig:Mass_Scatter} shows evolutionary tracks in the donor mass-period plane with the colour of each point representing the neutron star mass at that point. If a system experiences MT below the Eddington limit, the accretor mass can grow significantly. The significant increase in mass is a common outcome for default Skumanich MB, and the convection boosted cases, where most of the LMXBs evolve to contain a NS more massive than $2 M_\odot$. A ``stronger MB'', such as our intermediate case, produces fewer systems where NS masses have increased significantly, due to the MT exceeding the Eddington limit for portions of the evolution. The lack of substantial accretion of material onto the NS is even more apparent in the wind-boosted case where there is no significant increase in mass despite the large decrease in donor mass. However, the wind-boosted systems generally die too quickly -- most simulations don't reach the relevant M, P range for the majority of observed binaries. From observations there is a lack of NS detected near $2.0M_\odot$ with the most massive detected at $2.01 \pm 0.04 M_\odot$ \citep{Antoniadis2013}. The results shown in the default and convection boosted case suggest that all short period binaries contain a NS with a mass exceeding $2.0M_\odot$, whereas observations have found that these systems contain NS accretors in the range of $1.4M_\odot$. This preliminary result strongly supports the stronger MB prescriptions over the weaker convection boost and default cases.

\begin{figure*}
    \centering
    \includegraphics[width=0.45\textwidth]{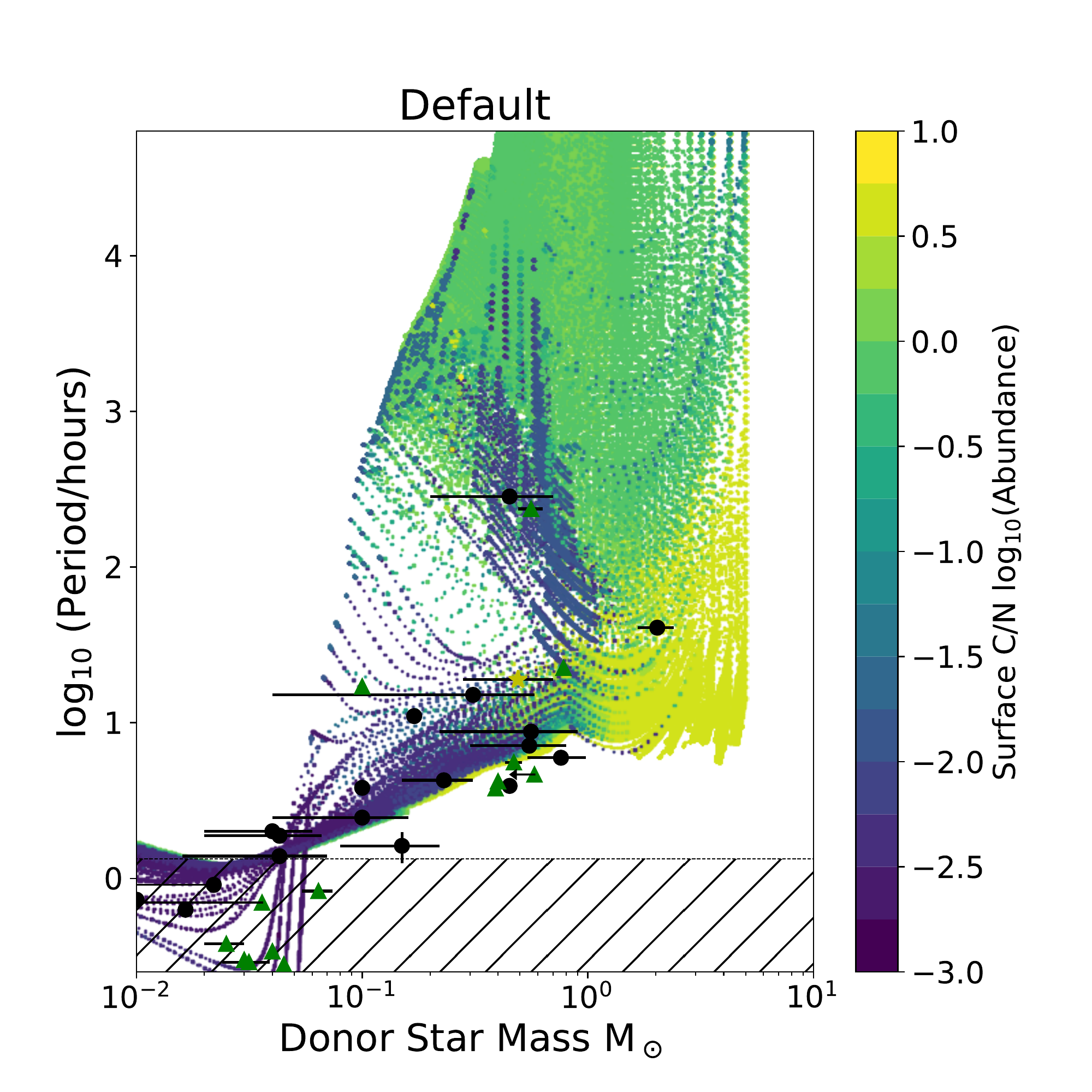}
    \hspace*{0.01cm}
    \includegraphics[width=0.45\textwidth]{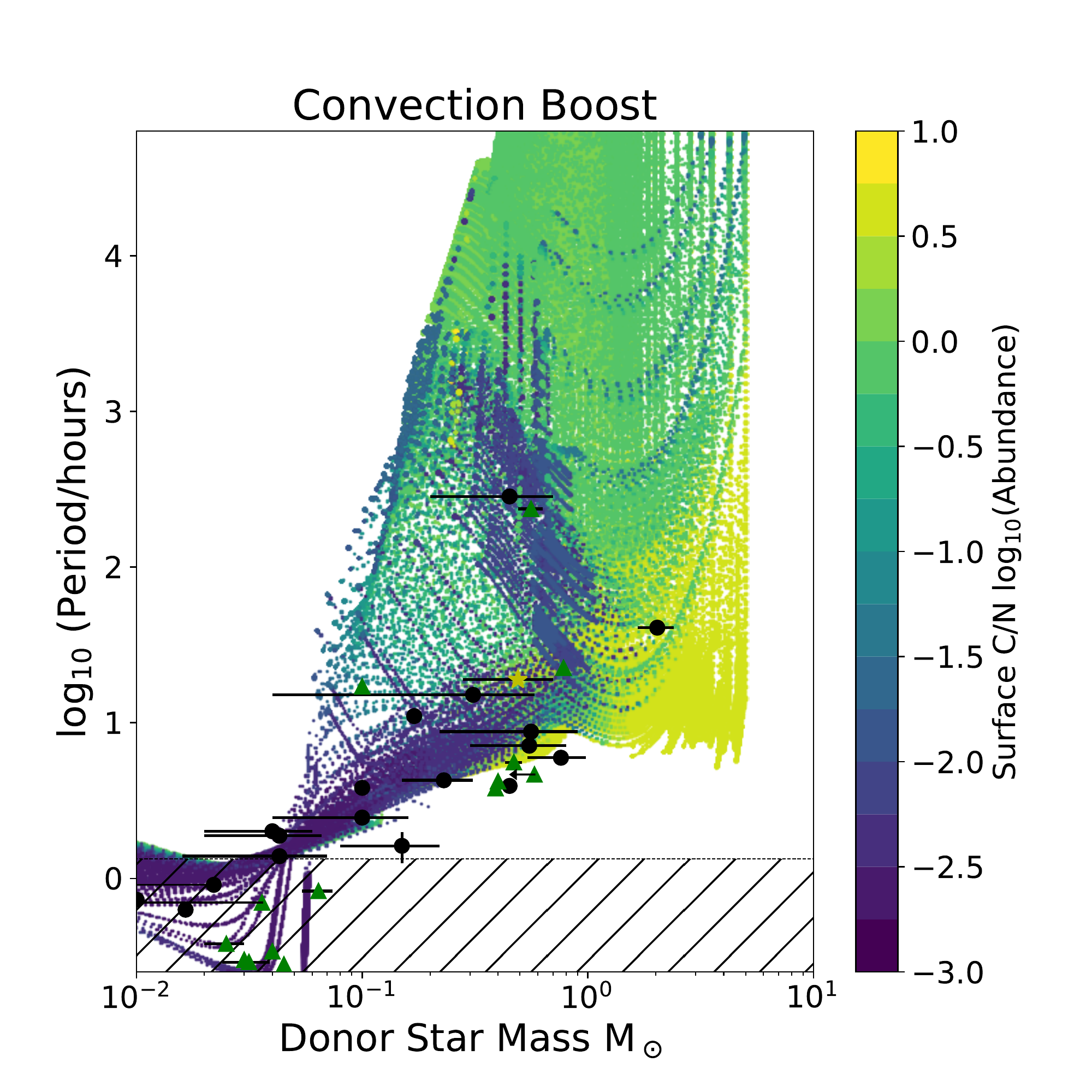}
    \includegraphics[width=0.45\textwidth]{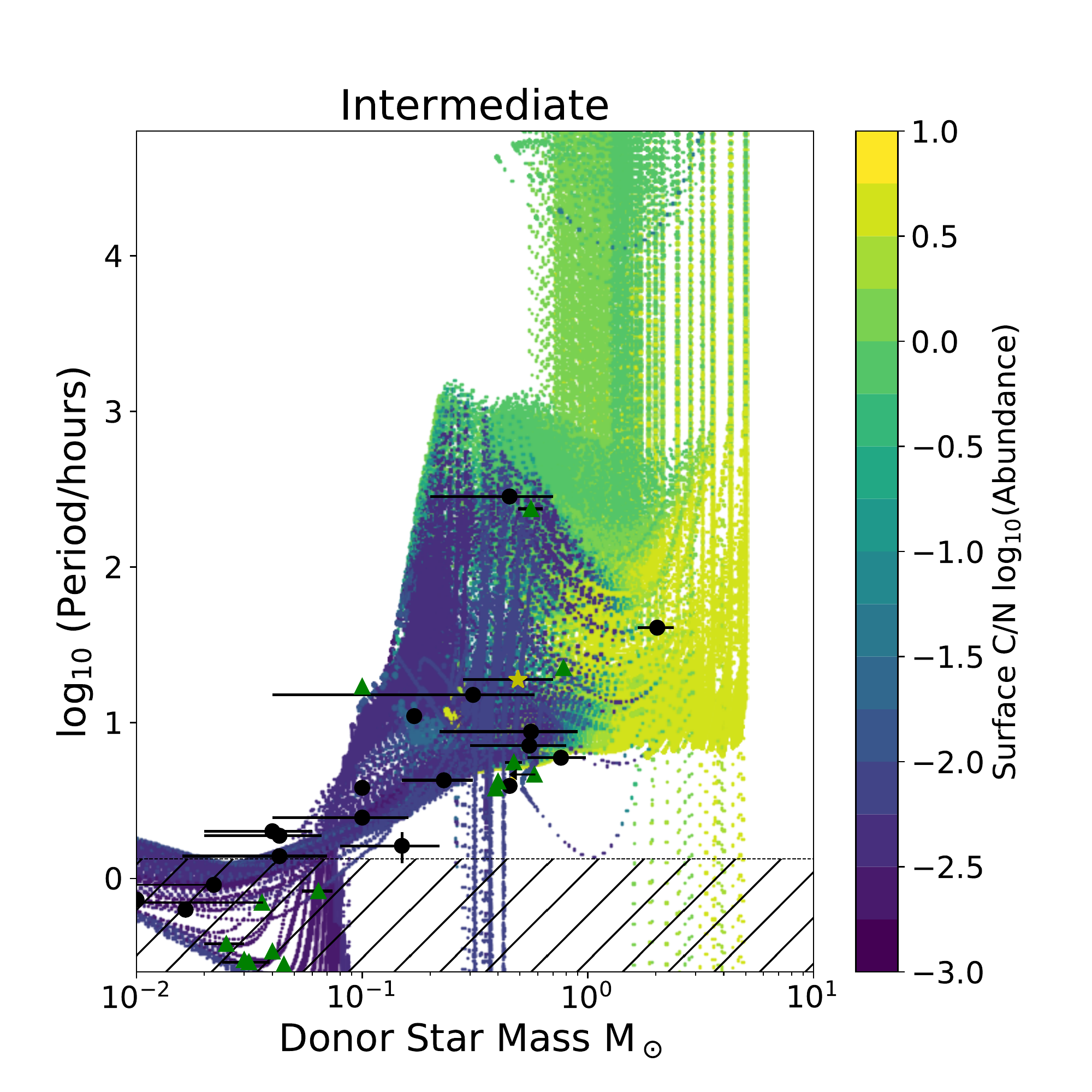}
    \hspace*{0.01cm}
    \includegraphics[width=0.45\textwidth]{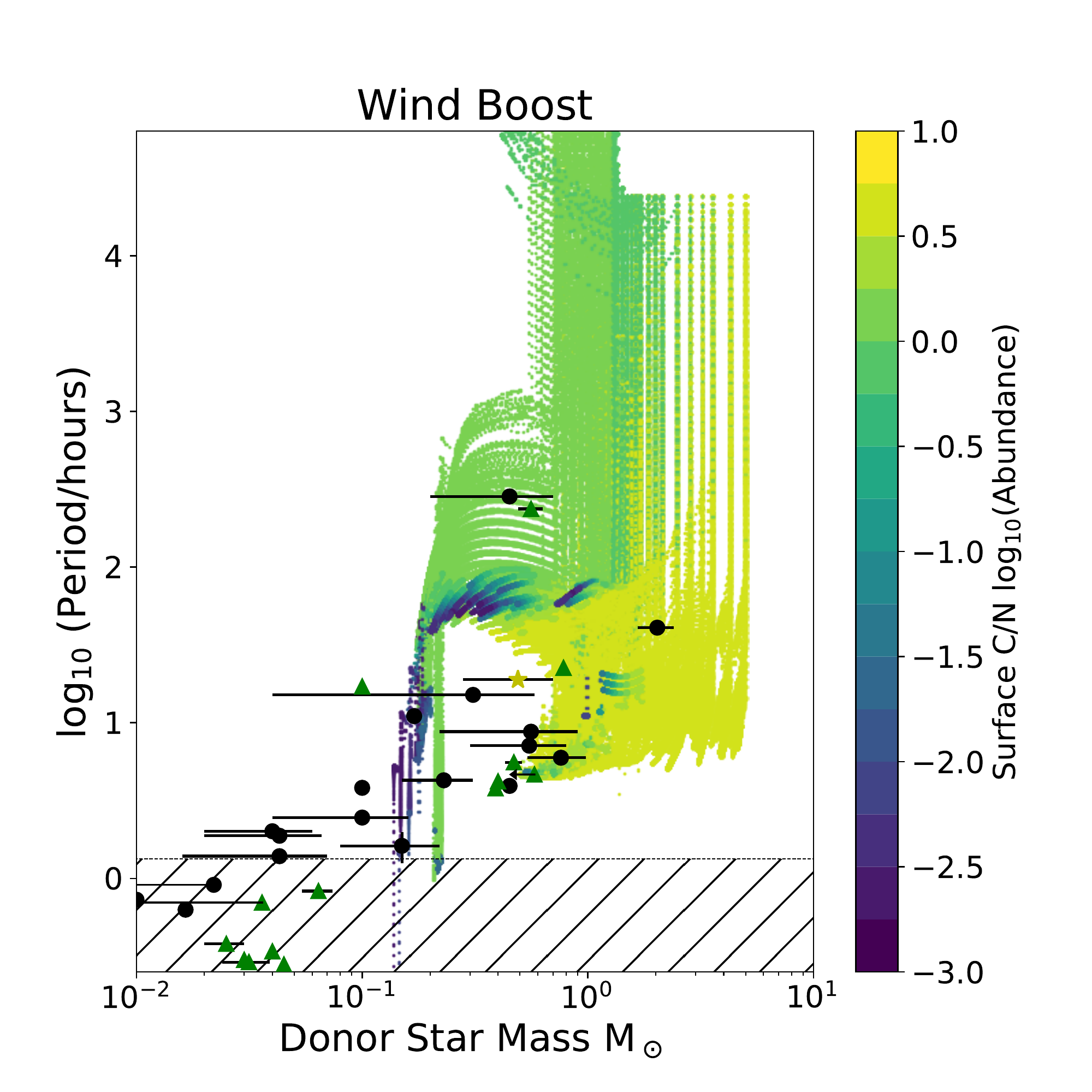}
    \caption{The surface C/N abundance mass fraction of the donor star. The initial C/N ratio of a star is $\sim$ 3.36 or $\sim$ 0.53 on the log scale.}
    \label{fig:Surf_CN_Scatter}
\end{figure*}

\begin{figure*}
    \centering
    \includegraphics[width=0.45\textwidth]{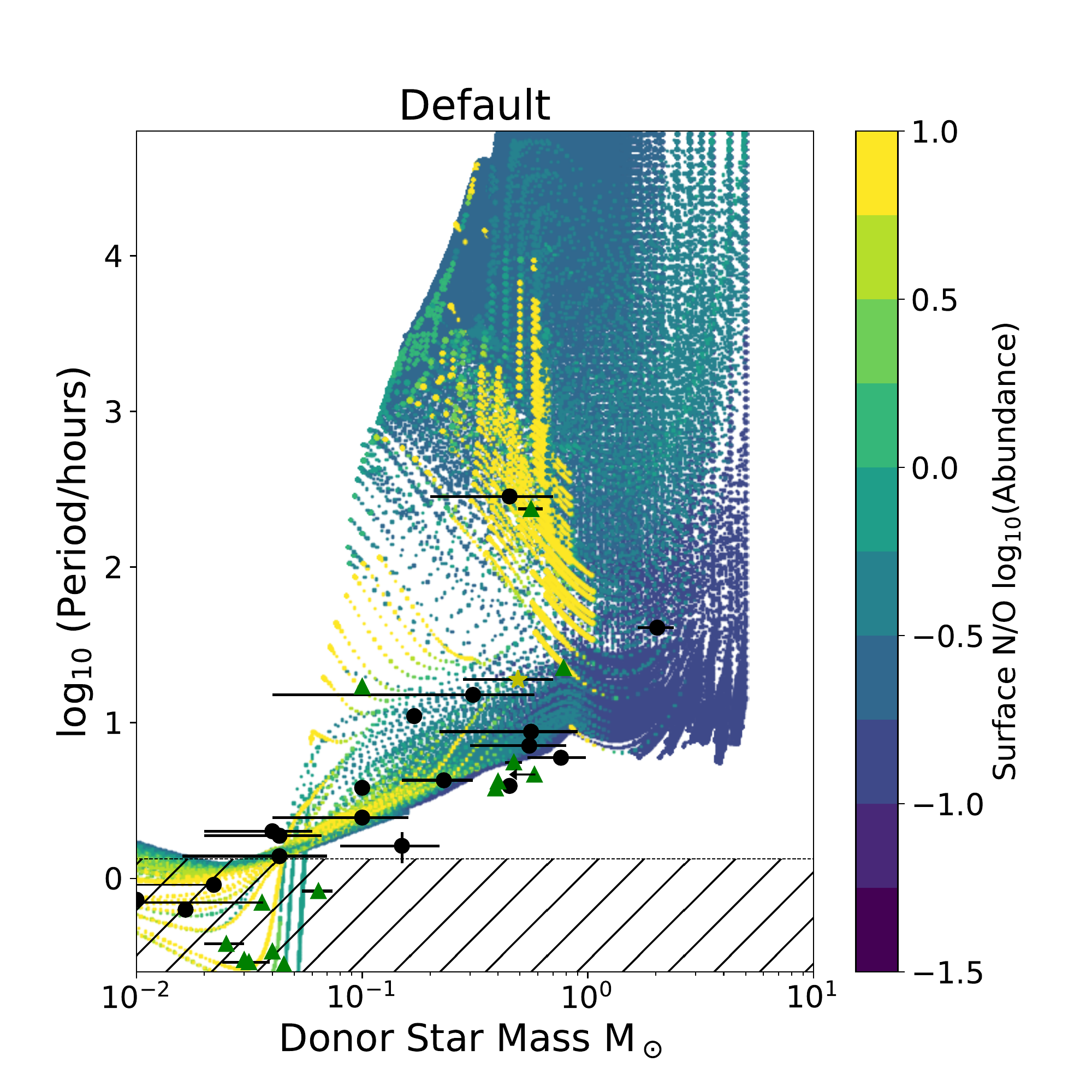}
    \hspace*{0.01cm}
    \includegraphics[width=0.45\textwidth]{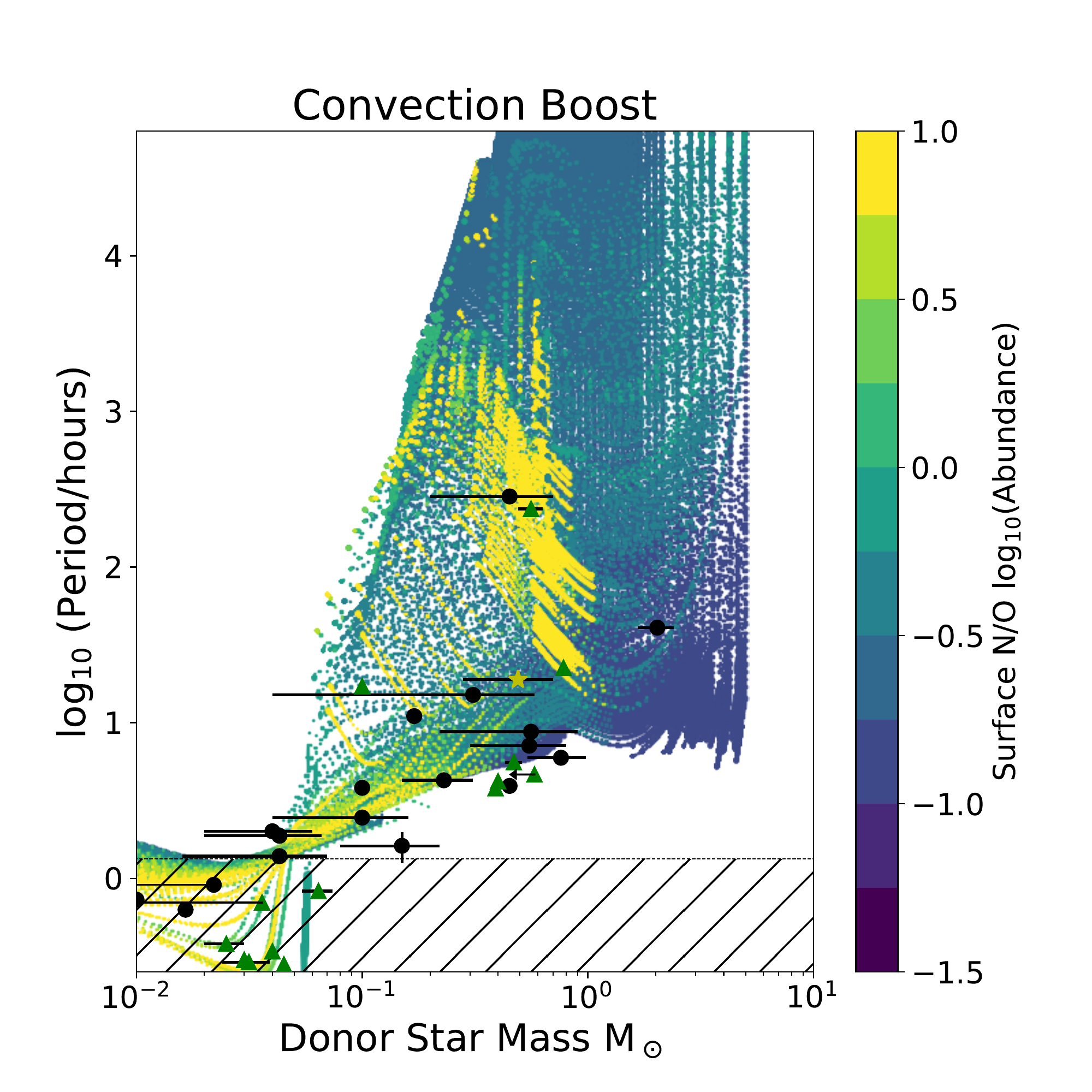}
    \includegraphics[width=0.45\textwidth]{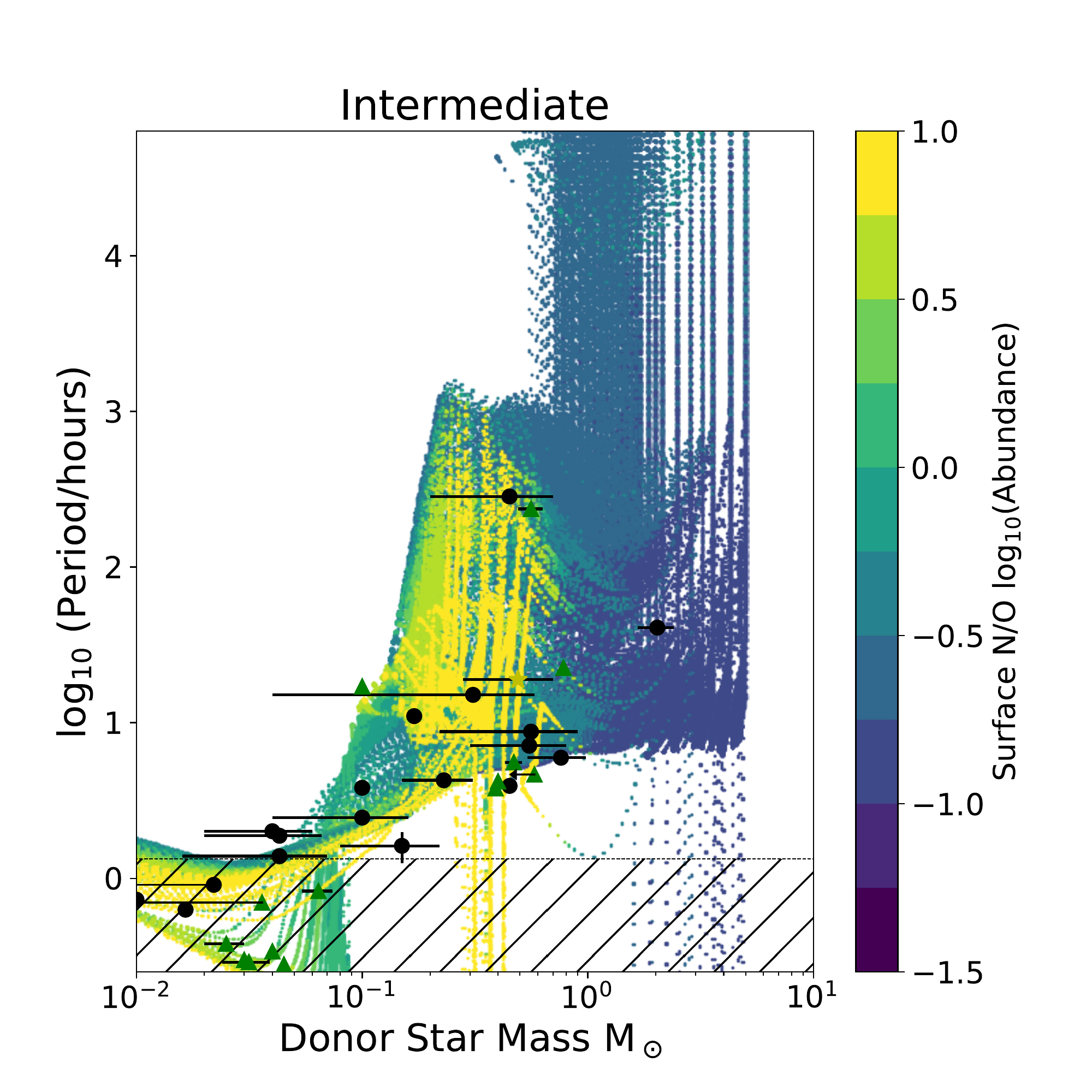}
    \hspace*{0.01cm}
    \includegraphics[width=0.45\textwidth]{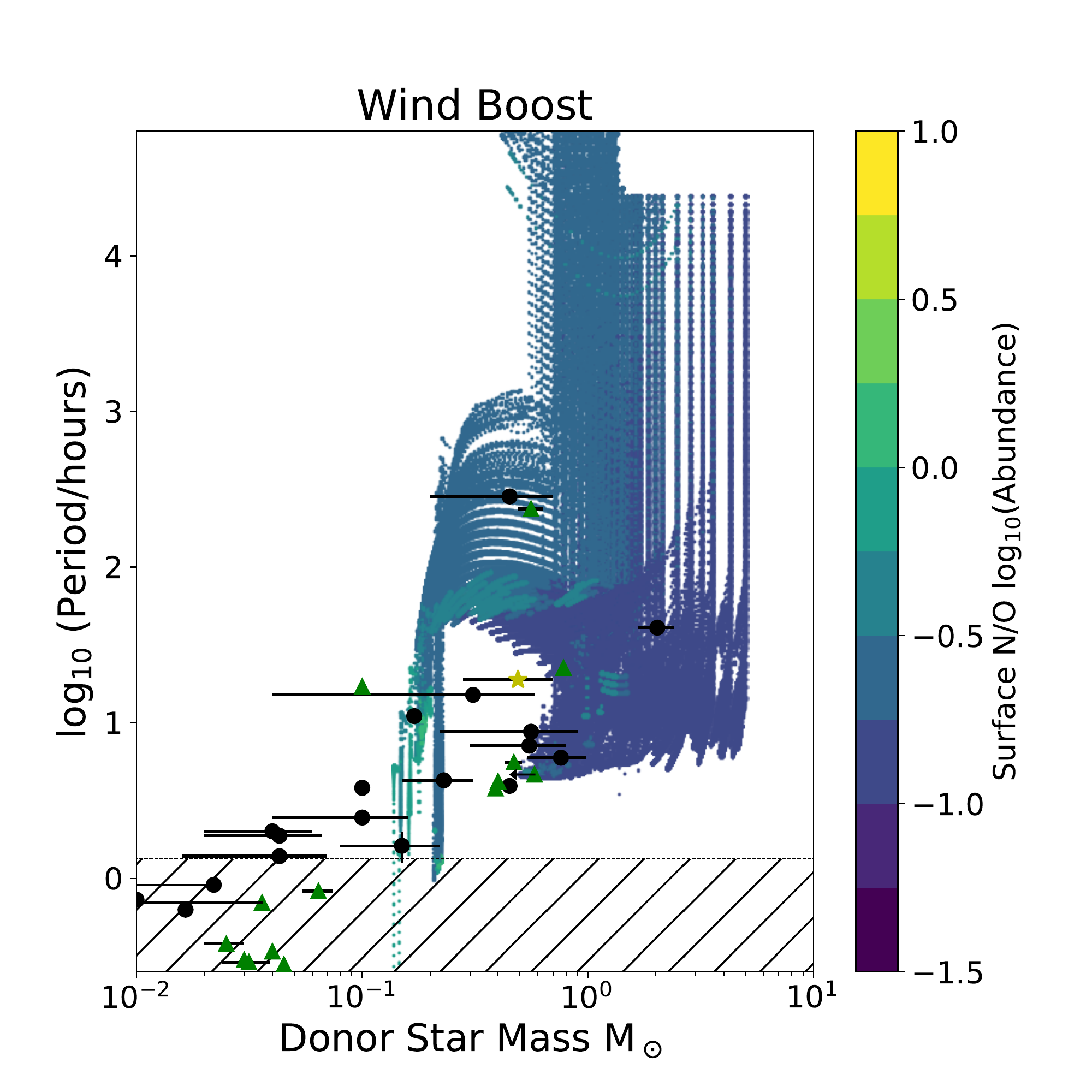}
    \caption{The surface N/O abundance mass fraction of the donor star. The initial N/O ratio of a star is 0.108 or $\sim$ -0.96 on the log scale.}
    \label{fig:Surf_NO_Scatter}
\end{figure*}

Measurement of the surface chemical composition is possible in select binary systems. In binary systems involving white dwarf accretors, the surface chemical abundances can constrain the possible formation channels, with high $\rm N/C > 10$ or $\rm N/O \gtrsim 10$ implying a helium donor \citep{Nelemans2010}. Uncertainties in model spectra for UCXBs result in unreliable abundance ratios \citep{Werner2006}. The existence of strong C and O lines but weak He and N lines imply a helium donor star. However, in many cases more detailed observations are necessary to classify the possible donors \citep{Nelemans2010}. Figures \ref{fig:Surf_CN_Scatter} and \ref{fig:Surf_NO_Scatter} show the C/N and N/O surface ratios from the simulated systems, respectively. In our donors, the initial C:N:O ratio is 0.37:0.108:1.0. 

This change of abundance indicates where the donor star is in its evolution. 
In systems with a lower-mass donor, $M \sim 1 M_\odot$, the donors show high N abundances as the donor star is stripped of C and O. In binaries with more massive donors, $M \gtrsim 1.5 M_\odot$, the CN and CNO cycles can result in substantial changes in chemical composition. These significant changes can be observed in the material that is being transferred in these compact binaries. 

\subsection{Relative Densities}
\label{subsec:Density}

\begin{figure*}
    \centering
    \includegraphics[width=0.45\textwidth]{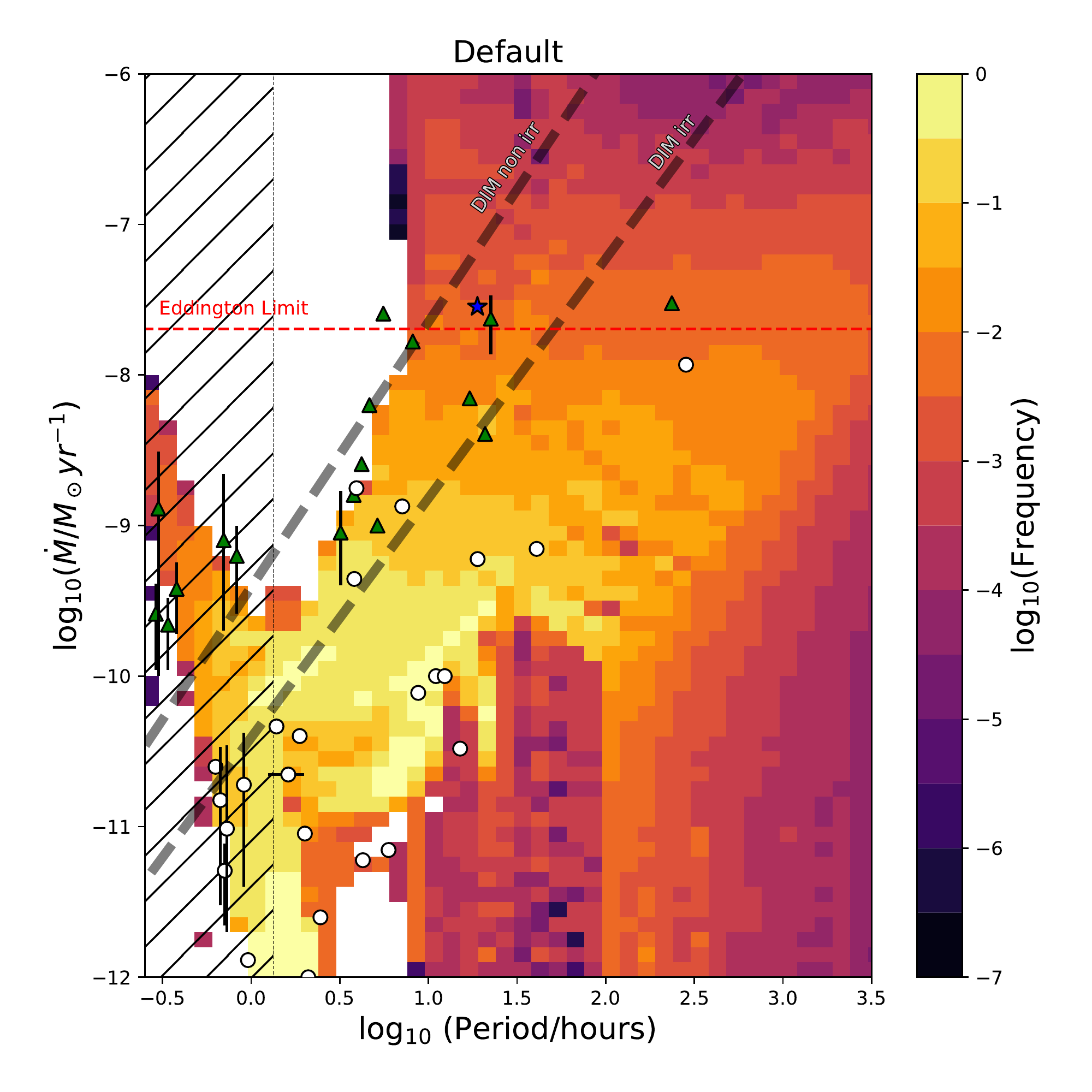}
    \hspace*{0.01cm}
    \includegraphics[width=0.45\textwidth]{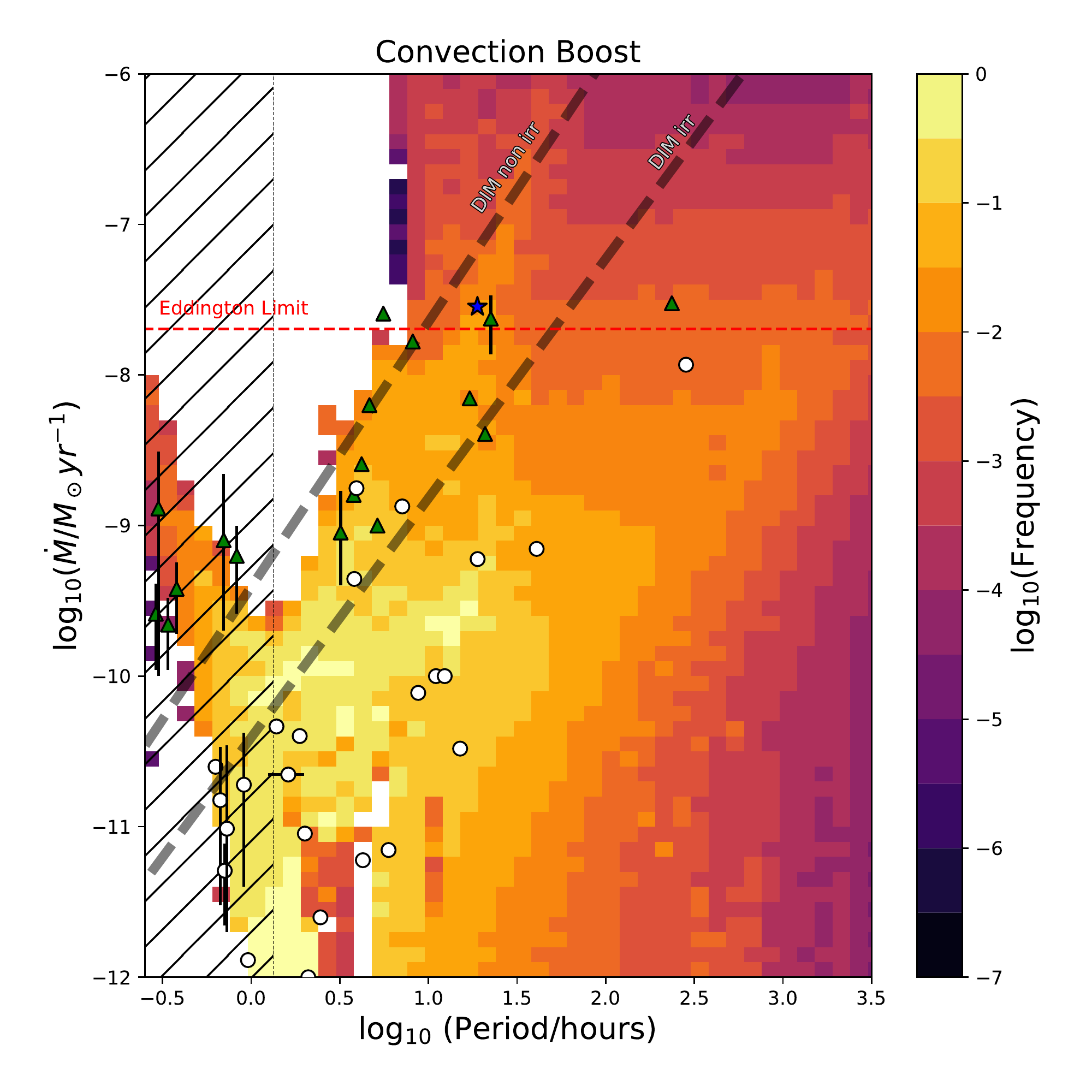}
    \includegraphics[width=0.45\textwidth]{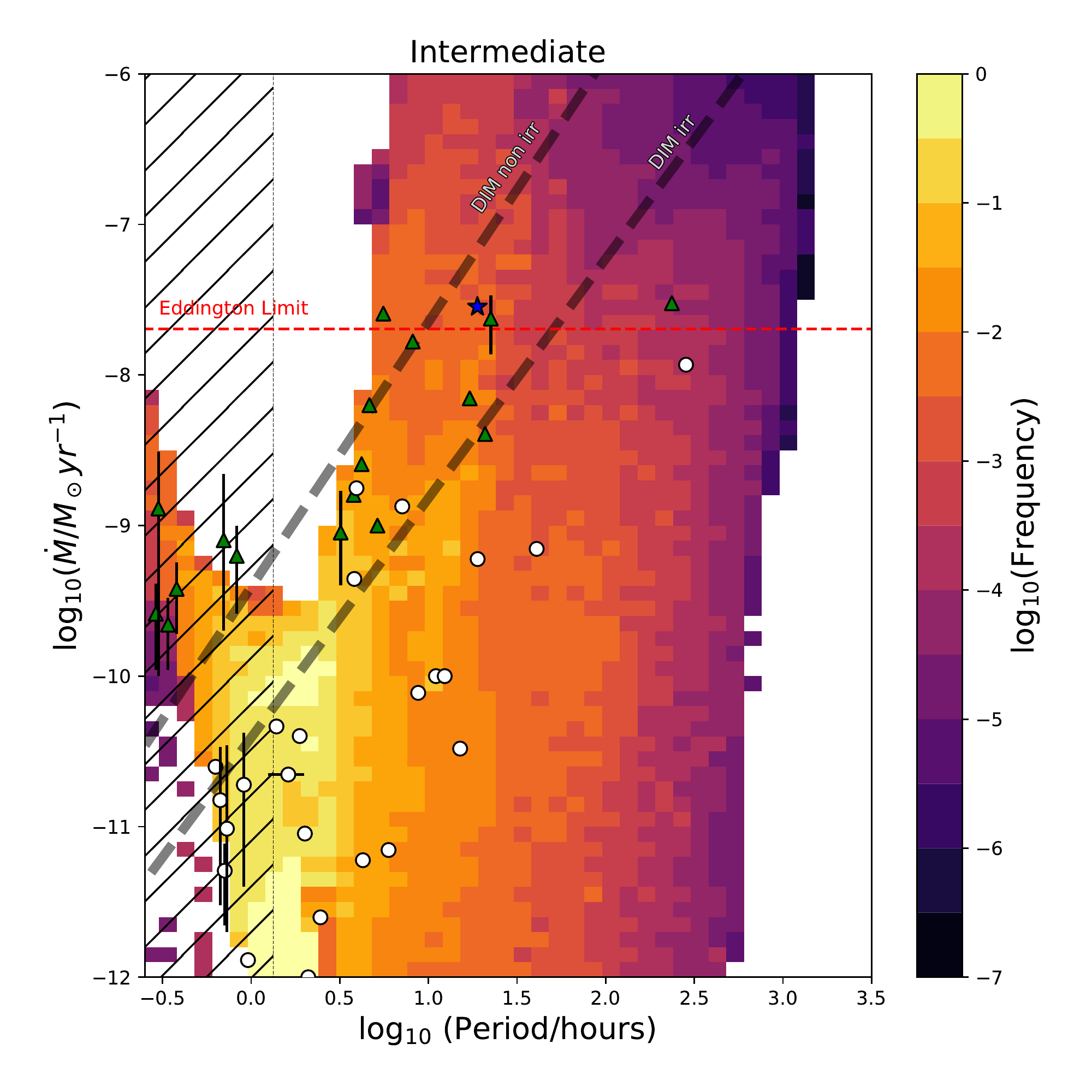}
    \hspace*{0.01cm}
    \includegraphics[width=0.45\textwidth]{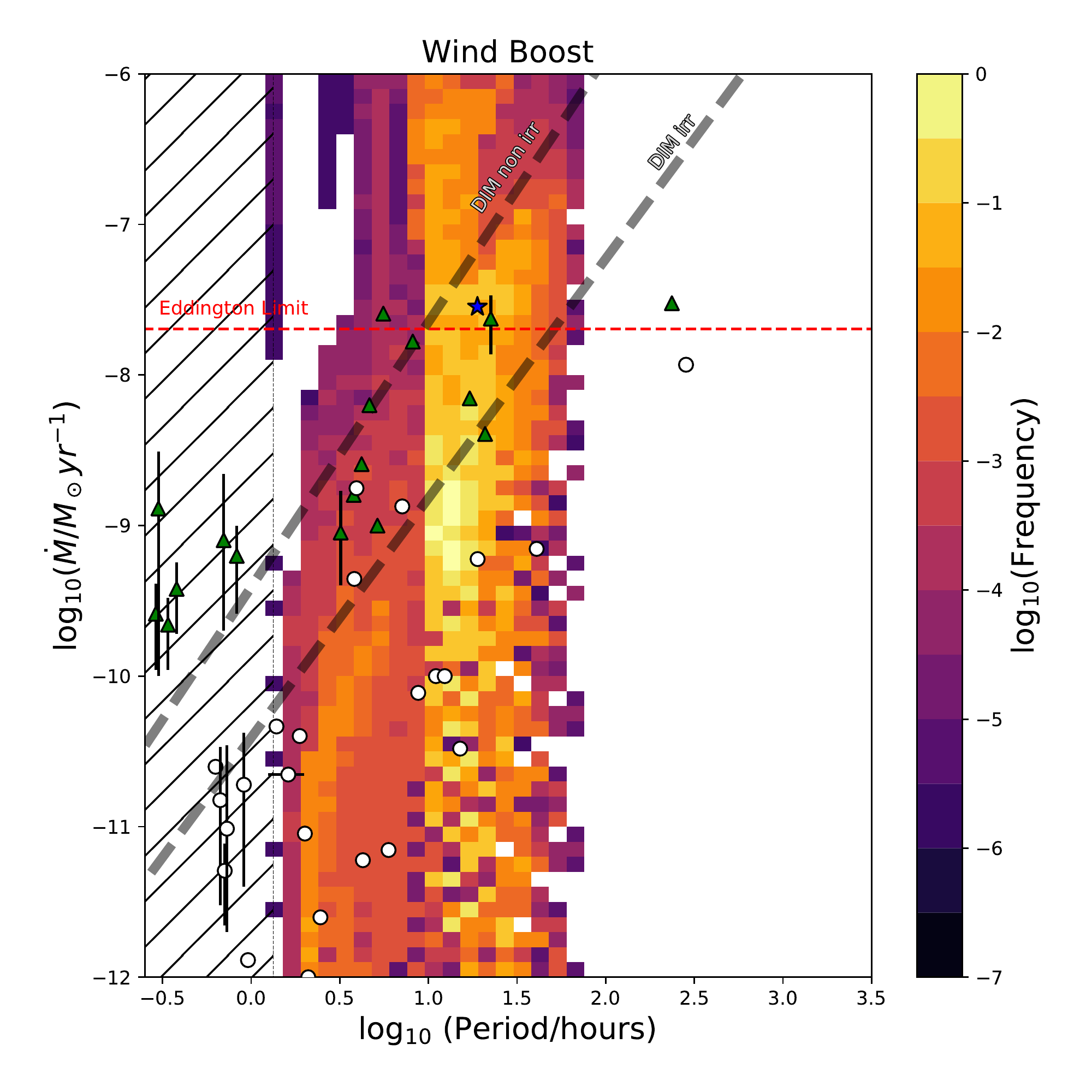}
    \caption{The relative probability of finding a system at a given point in our parameter space, assuming a the initial distribution of binaries from section \ref{sec:Binary_calculations}. The symbols on the plot represent the observed systems with their given errors (see Table \ref{NS_Table}). Circles are transient systems and triangles are persistent systems. The single star point is the binary Sco X-1. 
    The two grey dashed lines represent the critical MT rates for thermal-viscous accretion stability using an NS mass of 1.4$\rm M_\odot$ \citep{Coriat2012}. These two lines are denoted with ``DIM non irr'' for the critical MT rate without the effects of irradiation while ``DIM irr'' is the critical MT rate with the effects of irradiation. Binaries that lie above the line are predicted to be persistent LMXBs while systems below are transient. The left hashed area represents the period range for ultra-compact sources.
    }
    \label{fig:NS_Density}
\end{figure*}

Figure \ref{fig:NS_Density} shows the data plotted in the period-MT plane with the colour bar representing a normalized frequency. Here we can see which systems are more or less likely to appear in each respective bin. We calculate frequency as follows:

\begin{enumerate}
\item $\tau^{mn}_{\rm tot}$ is the total evolution time of a binary system, as defined by the initial $m$-period and $n$-mass.
\item $\tau_{ij}^{mn}$ is the time that a system defined by the initial $m$-period and $n$-mass spends in an $i$-period bin and a $j$-MT rate bin. 
\item $f_{ij}^{mn}=\tau_{ij}^{mn}/\tau^{mn}_{\rm tot}$  
is the fraction of time (or frequency) that a particular system (given by an initial $mn$) appears in a bin defined by a particular $i$-period and $j$-MT rate.
\item $f_{ij}=$ max $(f_{ij}^{mn})$ the frequency plotted in Fig. \ref{fig:NS_Density}, 
is the maximum frequency of all systems to appear in a particular $i$,$j$ bin. 
\end{enumerate}

The frequency gives an indication of where the evolutionary tracks spend the maximum amount of time during the evolution in the period-MT plane. This frequency however, does not necessarily represent a likelihood of detecting a binary in said bin as we have neglected effects of how likely a binary is to form and the number of systems which may cross through a bin. This is done to avoid equating regions where many systems cross into a bin for short periods of time to regions where one system spends a large fraction of its lifetime. For example, using this method, a bin where one binary spend $10^7$ years out of its total lifetime within the parameter space results in a much higher frequency than a bin where 1000 binaries each spend $10^4$ years.

We see the densest regions of the figure are at short periods, with low MT rates. These systems correspond to the binaries that have shrinking periods over the course of their evolution, including ultra-compact binaries. In general, the higher the MT rate, the lower the frequency; this is not surprising as it is difficult to maintain high MT rates. In the default MB, there is a gap in the simulated density between $ 0.5 \lesssim \log_{10}(\rm P/hr) \lesssim 1.0$ and $\log_{10}(\dot{M} / \Msun \rm yr^{-1}) \lesssim -11$. While there are no observed systems in this range, this gap begins to get populated once convection is accounted for and is filled in with the "intermediate" prescription. In general, as the MB boost is increased, the MT rate at lower periods is also increased. The "convection boost", "intermediate" and "wind boost" cases appear to more effectively reproduce the systems with periods $\log_{10}(\rm P/hr) \sim 1$ and MTs near the Eddington limit. These systems include GX 9+9, 4U 1735+444, and 2A 1822-371, which do not overlap with any simulated systems in figure \ref{fig:NS_Density}. Sco X-1, on the other hand, appears to be reproducible in figure \ref{fig:NS_Density}, as there is significant overlap with the simulated systems in the period-MT plane. We argue, however, that to reproduce an observed LMXB, in addition to the period and MT rate, we must also match the mass ratio of the system.

\section{Comparison to the observed population of LMXBs}
\label{sec:stat}

\begin{table}
\footnotesize
\caption{\textbf{Properties of Persistent LMXBs}}
\centering
\begin{tabular}{l | lll}

System Name          & $\log_{10}(P)$  & q            & $\log_{10}(\dot M_a)$ \\
\hline
\multicolumn{3}{l}{Ultra-compact XRBs } \tabularnewline 
4U 0513-40           & [-0.57, -0.52]  & [0.01, 0.06] & [-9.2, -8.6]     \\
2S 0918-549          & [-0.56, -0.51]  & [0.01, 0.06] & [-9.8, -8.6]     \\
4U 1543-624          & [-0.54, -0.49]  & [0.01, 0.06] & [-9.1, -8.6]     \\
4U 1850-087          & [-0.48, -0.43]  & [0.01, 0.06] & [-10.0, -8.4]     \\
M15 X-2              & [-0.44, -0.39]  & [0.01, 0.06] & [-9.7, -9.1]     \\
4U 1626-67           & [-0.17, -0.12]  & [0.01, 0.06] & [-9.7, -8.6]     \\
4U 1916-053          & [-0.10, -0.05]  & [0.03, 0.08] & [-9.6, -8.9]     \\
\hline 
\\
4U 1636-536          & [0.56, 0.61]    & [0.15, 0.40] & [-9.1, -8.6]     \\ 
GX 9+9               & [0.60, 0.65]    & [0.20, 0.33] & [-8.7, -8.2]     \\ 
4U 1735-444          & [0.65, 0.70]    & [0.29, 0.48] & [-8.4, -7.9]     \\ 
2A 1822-371          & [0.73, 0.78]    & [0.26, 0.36] & [-7.8, -7.3]     \\ 
Sco X-1              & [1.26, 1.31]    & [0.15, 0.58] & [-7.8, -7.3]     \\ 
GX 349+2             & [1.33, 1.38]    & [0.39, 0.65] & [-8.0, -7.3]     \\ 
Cyg X-2              & [2.35, 2.40]    & [0.25, 0.53] & [-7.7, -7.2]     \\  

\end{tabular}
\label{table:pers_table}
\begin{flushleft}
\textbf{Notes.} The binned properties of selected persistent NS LMXBs. The periods are in hours, the mass accretion rate $\dot M_a$ is in $M_\odot\ yr^{-1}$. The default bins' ranges are 0.05 in $\log_{10} P$ and 0.5 in $\log_{10} \dot M$, centred around the measured observed values. Ranges are increased if observational uncertainties are larger than the default ranges. The ranges for mass ratios, if those were not provided with an error, are such that they could accommodate the plausible error in NS mass, from 1.4 $M_\odot$ to the range in $1.2-2$ $M_\odot$.
\end{flushleft}
\end{table}

\begin{table}
\footnotesize
\caption{\textbf{Properties of Transient LMXBs}}
\centering
\begin{tabular}{l | lll}

System Name           & $\log_{10}(P)$  & q            & $\log_{10}(\dot M_a)$ \\ 
\hline

HETE J1900.1-2455     & [0.12, 0.17]    & [0.01, 0.06] & [-10.5, -10.0]   \\
1A 1744-361$^*$       & [0.19, 0.24]    & [0.05, 0.18] & [-11.7, -10.7]   \\
SAX J1808-3658        & [0.28, 0.33]    & [0.02, 0.07] & [-11.0, -10.5]   \\
IGR 00291+5394        & [0.37, 0.42]    & [0.02, 0.13] & [-11.8, -11.3]   \\
EXO 0748-676$^*$      & [0.56, 0.61]    & [0.05, 0.10] & [-9.3, -8.3]     \\
4U 1254-69            & [0.57, 0.62]    & [0.23, 0.38] & [-9.0, -8.5]     \\
XTE J1814-338$^*$     & [0.61, 0.66]    & [0.10, 0.27] & [-11.2, -10.2]   \\
XTE J2123-058$^*$     & [0.76, 0.81]    & [0.27, 0.82] & [-11.2, -10.2]   \\
X 1658-298            & [0.83, 0.88]    & [0.15, 0.67] & [-9.1, -8.6]     \\
SAX J1748.9-2021      & [0.92, 0.97]    & [0.06, 0.83] & [-10.3, -9.8]    \\
IGR J18245-2452       & [1.02, 1.07]    & [0.09, 0.14] & [-10.2, -9.7]    \\
Cen X-4               & [1.16, 1.21]    & [0.02, 0.48] & [-10.6, -10.1]   \\
Her X-1               & [1.59, 1.64]    & [0.83, 2.00] & [-8.1, -7.6]     \\
GRO J1744-28          & [2.43, 2.48]    & [0.25, 0.53] & [-8.2, -7.7]     \\

\end{tabular}
\label{table:trans_table}
\begin{flushleft}
\textbf{Notes.} The adopted ranges of selected transient NS LMXBs. Quantities are as in Table~\ref{table:pers_table} except in systems with a * symbol. The mass transfer bins in these systems use the upper limit listed, and span $\log_{10} \dot M = 1.0$. 
\end{flushleft}
\end{table}

\begin{table*}

\caption{\textbf{Maximum Lifetime of an Observed LMXB and Fraction of Parameter Space}}
\centering
\begin{tabular}{l | cc | cc | cc | cc }

                   & \multicolumn{2}{c|}{Default} & \multicolumn{2}{c|}{Convection Boosted} & \multicolumn{2}{c|}{Intermediate} & \multicolumn{2}{c|}{Wind Boost} \\
System Name        & $\tau_{\rm max}$ [years] & $A_{\rm sys}/A_{\rm tot}$ & $\tau_{\rm max}$ [years] & $A_{\rm sys}/A_{\rm tot}$ & $\tau_{\rm max}$ [years] & $A_{\rm sys}/A_{\rm tot}$ & $\tau_{\rm max}$ [years] & $A_{\rm sys}/A_{\rm tot}$ \\
\hline
4U 0513-40         & $ 3.23\times10^{6} $ & $ 4.75\times10^{-4} $ & $ 1.24\times10^{6} $ & $ 7.91\times10^{-4} $ & $ 4.00\times10^{6} $ & $ 1.90\times10^{-3} $ & $ 0                $ & $ 0                 $ \\ 
2S 0918-549        & $ 4.09\times10^{6} $ & $ 4.75\times10^{-4} $ & $ 2.30\times10^{6} $ & $ 7.91\times10^{-4} $ & $ 4.40\times10^{6} $ & $ 2.06\times10^{-3} $ & $ 0                $ & $ 0                 $ \\ 
4U 1543-624        & $ 4.83\times10^{6} $ & $ 6.33\times10^{-4} $ & $ 4.80\times10^{6} $ & $ 6.33\times10^{-4} $ & $ 7.91\times10^{6} $ & $ 2.06\times10^{-3} $ & $ 0                $ & $ 0                 $ \\ 
4U 1850-087        & $ 8.33\times10^{6} $ & $ 6.33\times10^{-4} $ & $ 8.83\times10^{6} $ & $ 1.11\times10^{-3} $ & $ 1.28\times10^{7} $ & $ 2.06\times10^{-3} $ & $ 0                $ & $ 0                 $ \\ 
M15 X-2            & $ 1.21\times10^{7} $ & $ 6.33\times10^{-4} $ & $ 1.11\times10^{7} $ & $ 1.27\times10^{-3} $ & $ 1.41\times10^{7} $ & $ 2.22\times10^{-3} $ & $ 0                $ & $ 0                 $ \\ 
4U 1626-67         & $ 6.76\times10^{7} $ & $ 9.50\times10^{-4} $ & $ 1.06\times10^{8} $ & $ 1.58\times10^{-3} $ & $ 7.48\times10^{7} $ & $ 6.33\times10^{-4} $ & $ 0                $ & $ 0                 $ \\ 
4U 1916-053        & $ 1.85\times10^{7} $ & $ 1.58\times10^{-4} $ & $ 1.21\times10^{7} $ & $ 3.17\times10^{-4} $ & $ 3.76\times10^{7} $ & $ 4.75\times10^{-4} $ & $ 0                $ & $ 0                 $ \\ 
\hline
4U 1636-536        & $ 1.22\times10^{8} $ & $ 4.23\times10^{-2} $ & $ 6.24\times10^{7} $ & $ 1.91\times10^{-2} $ & $ 4.39\times10^{7} $ & $ 7.74\times10^{-2} $ & $ 1.22\times10^{5} $ & $ 2.06\times10^{-3} $ \\ 
GX 9+9             & $ 0                $ & $ 0                 $ & $ 3.38\times10^{7} $ & $ 1.60\times10^{-2} $ & $ 3.73\times10^{7} $ & $ 5.89\times10^{-2} $ & $ 1.85\times10^{4} $ & $ 1.27\times10^{-3} $ \\ 
4U 1735-444        & $ 0                $ & $ 0                 $ & $ 1.48\times10^{7} $ & $ 5.70\times10^{-3} $ & $ 9.83\times10^{6} $ & $ 2.26\times10^{-2} $ & $ 0                $ & $ 0                 $ \\ 
2A 1822-371        & $ 0                $ & $ 0                 $ & $ 0                $ & $ 0                 $ & $ 5.26\times10^{6} $ & $ 3.39\times10^{-2} $ & $ 0                $ & $ 0                 $ \\ 
Sco X-1            & $ 0                $ & $ 0                 $ & $ 1.69\times10^{7} $ & $ 3.48\times10^{-3} $ & $ 1.75\times10^{6} $ & $ 3.91\times10^{-2} $ & $ 2.71\times10^{3} $ & $ 2.37\times10^{-3} $ \\ 
GX 349+2           & $ 4.09\times10^{5} $ & $ 7.91\times10^{-4} $ & $ 1.50\times10^{7} $ & $ 6.17\times10^{-3} $ & $ 1.91\times10^{5} $ & $ 2.37\times10^{-3} $ & $ 0                $ & $ 0                 $ \\ 
Cyg X-2            & $ 1.65\times10^{6} $ & $ 8.39\times10^{-3} $ & $ 2.56\times10^{6} $ & $ 1.74\times10^{-2} $ & $ 2.00\times10^{5} $ & $ 1.28\times10^{-2} $ & $ 0                $ & $ 0                 $ \\ 
\hline
HETE J1900.1-2455  & $ 5.61\times10^{8} $ & $ 6.61\times10^{-2} $ & $ 3.05\times10^{8} $ & $ 4.57\times10^{-2} $ & $ 5.87\times10^{8} $ & $ 1.01\times10^{-1} $ & $ 0                $ & $ 0                 $ \\
1A 1744-361$^*$    & $ 1.99\times10^{6} $ & $ 1.58\times10^{-4} $ & $ 3.51\times10^{7} $ & $ 1.58\times10^{-4} $ & $ 6.80\times10^{7} $ & $ 7.91\times10^{-4} $ & $ 0                $ & $ 0                 $ \\
SAX J1808-3658     & $ 5.49\times10^{7} $ & $ 6.33\times10^{-4} $ & $ 1.05\times10^{8} $ & $ 6.33\times10^{-4} $ & $ 1.66\times10^{8} $ & $ 1.74\times10^{-3} $ & $ 0                $ & $ 0                 $ \\
IGR 00291+5394     & $ 0                $ & $ 0                 $ & $ 0                $ & $ 0                 $ & $ 1.88\times10^{8} $ & $ 4.75\times10^{-4} $ & $ 0                $ & $ 0                 $ \\
EXO 0748-676$^*$   & $ 0                $ & $ 0                 $ & $ 4.09\times10^{7} $ & $ 2.53\times10^{-3} $ & $ 0                $ & $ 0                 $ & $ 0                $ & $ 0                 $ \\
4U 1254-69         & $ 5.23\times10^{7} $ & $ 1.28\times10^{-2} $ & $ 1.06\times10^{6} $ & $ 7.91\times10^{-4} $ & $ 1.70\times10^{7} $ & $ 1.65\times10^{-2} $ & $ 0                $ & $ 0                 $ \\
XTE J1814-338$^*$  & $ 0                $ & $ 0                 $ & $ 0                $ & $ 0                 $ & $ 0                $ & $ 0                 $ & $ 2.60\times10^{5} $ & $ 1.27\times10^{-3} $ \\
XTE J2123-058$^*$  & $ 0                $ & $ 0                 $ & $ 0                $ & $ 0                 $ & $ 1.40\times10^{3} $ & $ 1.58\times10^{-3} $ & $ 2.25\times10^{2} $ & $ 2.53\times10^{-3} $ \\
X 1658-298         & $ 1.19\times10^{8} $ & $ 1.30\times10^{-2} $ & $ 2.85\times10^{7} $ & $ 8.23\times10^{-3} $ & $ 1.55\times10^{7} $ & $ 4.91\times10^{-3} $ & $ 2.26\times10^{4} $ & $ 3.17\times10^{-3} $ \\
SAX J1748.9-2021   & $ 7.55\times10^{8} $ & $ 6.33\times10^{-4} $ & $ 3.08\times10^{8} $ & $ 1.27\times10^{-3} $ & $ 4.92\times10^{7} $ & $ 1.39\times10^{-2} $ & $ 8.96\times10^{4} $ & $ 2.69\times10^{-3} $ \\
IGR J18245-2452    & $ 8.91\times10^{8} $ & $ 6.33\times10^{-4} $ & $ 0                $ & $ 0                 $ & $ 3.82\times10^{7} $ & $ 2.37\times10^{-3} $ & $ 0                $ & $ 0                 $ \\
Cen X-4            & $ 0                $ & $ 0                 $ & $ 1.14\times10^{8} $ & $ 1.58\times10^{-4} $ & $ 3.53\times10^{7} $ & $ 1.58\times10^{-2} $ & $ 4.32\times10^{3} $ & $ 1.11\times10^{-3} $ \\
Her X-1            & $ 7.38\times10^{6} $ & $ 1.27\times10^{-3} $ & $ 9.97\times10^{6} $ & $ 1.27\times10^{-3} $ & $ 5.68\times10^{5} $ & $ 1.42\times10^{-3} $ & $ 4.75\times10^{5} $ & $ 2.69\times10^{-3} $ \\
GRO J1744-28       & $ 5.85\times10^{6} $ & $ 8.70\times10^{-3} $ & $ 5.48\times10^{6} $ & $ 1.31\times10^{-2} $ & $ 6.98\times10^{4} $ & $ 5.70\times10^{-3} $ & $ 0                $ & $ 0                 $ \\
\end{tabular}
\label{table:res_table}
\begin{flushleft}
\textbf{Notes.} For each observed system (and thus parameter space bin), we give the maximum amount of time, $\tau_{\rm max}$, that any simulated system spends in a given bin, and the fraction of the parameter space of simulated binaries which resemble the observed system, $A_{\rm sys}/A_{\rm tot}$. Our bins are defined in tables \ref{table:pers_table} and \ref{table:trans_table}. The total parameter space $A_{\rm tot}$ spans the mass range from $1 \leq M/M_\odot \leq 7$ and the period range from $-0.5 \leq \log_{10}(\rm P/days) \leq 4 $. The $*$ denotes systems where only an upper limit for the mass transfer rate is given. 
\end{flushleft}
\end{table*}

We describe 
a set of 
binary systems that have an observationally determined MT rate, orbital period, and  mass ratio (see Table ~\ref{NS_Table}, not all observed systems can be used). We bin each binary system within a range of period, MT and mass ratio, with the observed values used as the central bin values. These bins are then used to analyze the systems, and the adopted range for $P$, $\dot M$ and $q$ are described in Tables~\ref{table:pers_table} and \ref{table:trans_table}. The ``observational'' bin sizes are large enough to accommodate the anticipated observational errors. 

The likelihood of a given MB scheme being correct depends on how effectively it can reproduce the observed binary systems. To find this value, we check if a simulated MT system passes through any of the ``observational'' bins. If a simulated system passes through one of the observational bins, we can find the total time that the system spent $\tau_{ij}^n$ in that ``observational'' bin. Here, $n$ stands for the number of the ``observational'' bin, and $ij$ describes the initial orbital period and the initial mass of the donor. We then find the size of the initial parameter space in the initial orbital periods and the donor masses, the parameter space from which the systems could evolve through the particular ``observational'' bin $n$.

In Table \ref{table:res_table} we provide the maximum amount of time, $\tau_{\rm max}^{n}$ that any of the simulated systems can spend in the $n$ observational bin of interest, and the fraction of the initial parameter space that can produce the observed systems. These results are separated by MB prescription, with the wind-boosted case producing clearly ineffective results. With the high MT rate, the simulated parameter space overlaps with very few observed data points in \ref{fig:WB_mass_grid} and \ref{fig:NS_Density}. This high boost rate is likely invalid, and including a dampening factor should yield more realistic results. 

The most striking result is that the default Skumanich MB law cannot reproduce most of the persistent systems - in fact, no observed persistent LMXBs with orbital periods between about 4.5 and 23 hours can be produced once we account for mass ratio. The main reason is that MB is not strong enough to drive the observed MT rate. In principle, in addition to convection or wind-boosted MB laws, one can invoke also another alternative MB law to explain systems like Sco X-1 - as was done, e.g., by \cite{Chen2017}, who considered initial donors of $1.5-2.5 M_\odot$ star and applied the MB law derived from \cite{Justham2006}. The best fitting progenitor systems from \cite{Chen2017} have $1.6-1.8 M_\odot$ donors with a 300G fossil magnetic field. It would, however, be rather intriguing if most of the observed persistent systems must be descendants of low-mass A stars with magnetic fields about an order of magnitude weaker than that of Ap stars, but 100 times larger than that of regular stars.

A significant result is that while we can create UCXBs, the initial parameter space to form these systems, and the lifetimes of the systems in those data bins, suggest that binary systems with a NS accretor and a non-perturbed donor are unlikely to be the main progenitors. They are instead likely to be produced either in globular clusters via physical collisions of a NS with a red giant \cite{Ivanova2005}, or as a result of common evolution in the field, where the MT can start either from a cooled-down stripped core or, conversely, from a hot stripped core \citep{Heinke2013}. While the UCXBs have low MT rates, the short time spent in the observed bin is due to the simulated binary having an MT rate near the boundary of the bin.

As the MB strength increases, the number of persistent systems that can be reproduced increases. The binaries that could not be reproduced by the default MB prescription, GX 9+9, 4U 1735-444 and Sco X-1, are those that \cite{Podsiadlowski2002} found had MT rates much higher than their simulations reached.  
The convection-boosted MB is necessary to reproduce these persistent systems. Similarly, the available evidence indicates that the MT rate of 2A 1822-371 is super-Eddington \citep{Bak2017}. For 2A 1822-371, the convection-boosted case is still insufficient to reach the high MT observed. The wind-boosted case pushes the MT rate high enough to reproduce this system at an appropriate period. The mass ratio however, does not match with the observed binary.

The reproducibility of a transient binary is affected by the choice of MB prescription. Unfortunately, unlike persistent systems, a clear trend isn't apparent from one prescription to another. Instead, different systems are reproduced by different MB prescriptions, and with the uncertainties in determining an average MT in these systems, we cannot use the transient binaries to reliably draw any conclusions without a clear trend. One result that can be seen from the reproducibility of the transient systems in Table \ref{table:res_table} is that the systems that are most difficult to reproduce are those where the MT rate is only constrained by an upper limit. The intermediate case reproduces the largest number of observed transient systems with only EXO 0748-676 and XTE J1814-338 not being reproduced.

The wind-boosted MB simulations 
cannot reproduce the majority of the observed binaries regardless of whether they are persistent or transient. The wind-boost prescription 
gives very short lifetimes for all reproduced binaries (which makes their detection unlikely), and cannot reproduce the UCXBs. This suggests that the simulated wind-boosted case is exceeding some saturation point for MB and the systems are losing too much angular momentum too quickly.

\subsection{Other effects}
Irradiation may play a significant role in driving winds from the donor star \citep{Ruderman1989}, and in causing the donor star to expand to a larger radius than expected for its orbital period \citep{Podsiadlowski1991}. Such an increase in donor radius due to irradiation may cause cycles of increased MT rates (up to a factor of $\sim$30), followed by a decrease to below-average MT rates \citep[e.g.][]{Hameury1993}. The detailed physics of such irradiation-driven MT cycles have not yet been established, but current work suggests that these cycles should require small convective timescales in the donor, and thus may operate on systems with periods between 4 and 15 hours \citep{Buning2004}.  Irradiation-induced MT cycles could potentially produce the large observed MT rates in some of our transient and persistent systems in this period range, but should not be relevant for longer-period systems such as Sco X-1. Whether irradiation-induced MT can play a significant role depends on as-yet-undetermined details of the heating efficiency of the irradiating flux, and the fraction of time at the increased MT rate.

We have difficulty reproducing two transient systems (EXO 0748-676 and XTE J1814-338), both with orbital period near 4 hours, with any of our MB schemes. Our best MB scheme, the "intermediate" case, predicts MT rates higher than observed for these 2 systems. It is possible that irradiation-driven MT cycles might alter these systems' evolution enough to match their observed characteristics (although such MT cycles are thought to cut off around 4 hours, \citealt{Buning2004}). An alternative possibility is that these systems may turn on as millisecond radio pulsars intermittently, during which they eject all mass transferred from their companion \citep{Burderi2002,Burderi2003}, as the transitional millisecond pulsars appear to do  \citep{Archibald09,Papitto2013}. If so, the time-averaged MT rate onto their NSs would be lower than we calculate. There is indeed evidence that these two systems may be transitional millisecond pulsars. XTE J1814-338 is known to show accretion-induced X-ray pulsations \citep{Markwardt03b}.  Its donor star also shows evidence of irradiation by an unknown energy source, which may be spin-down energy from a radio pulsar \citep{Baglio13,Wang2017}. EXO 0748-676 has not shown detected X-ray pulsations (despite sensitive RXTE X-ray observations). However, a careful study during quiescence showed that no accretion disk was present, which may indicate that a transitional millisecond pulsar had turned on, and is ejecting transferred mass \citep{Ratti2012}.


\section{Conclusions}

In this work, we have examined how different MB prescriptions affect the evolution of LMXBs. The observational data to which we compare our simulations is given in section \ref{sec:Observational Data}. By systematically studying the parameter space of interest, we cover a range of possible seed masses and periods for these binary systems. The results of 
comparing our simulations to the observations 
are given in sections \ref{sec:Principle Results} and \ref{sec:stat}. The key results of this work are:

\begin{itemize}
    \item Using "weaker" MB schemes such as the default Skumanich prescription, and even the "convection boosted" case described in this work, results in an overabundance of highly massive NS accretors. 
    \item The highest density region in our parameter space, as seen in figure \ref{fig:NS_Density}, is the region of short periods and low MT rates. This high-density region is found in all MB prescriptions.
    \item In the default, convection boosted and intermedaite MB prescriptions, all UXCBs of interest can be reproduced. Although, $\tau_{\rm max}$ is small suggesting these systems are difficult to form using this method.
    \item The "default" MB scheme reproduces results similar to \cite{Podsiadlowski2002}. 
    This weak 
    MB scheme cannot reproduce some observed persistent systems in our simulations; these simulated binaries differ from observed binaries by up to an order of magnitude in MT rate.
    \item The "convection-boosted" prescription reproduces persistent systems much better than the default scheme, as it successfully simulates the properties of GX 9+9, 4U 1735-444 and Sco X-1. It cannot, however, reproduce the suspected super-Eddington system 2A 1822-371 \citep{Bak2017}.
    \item Once we account for wind in the MB scheme, we can reach high enough MT rates to reproduce 2A 1822-371.  Super-Eddington MT rates are achieved in the "intermediate" MB prescription.
    \item The intermediate prescription produces the largest number of observed transient LMXBs. Only EXO 0748-676 and XTE J1814-338 cannot be reproduced. These two systems only have an upper limit for mass transfer rate which my be the reason why these systems are difficult to reproduce.
    \item Including the effects of a non-thermal wind in our "wind boost" case results in very high MB. The high angular momentum loss results in MT rates that exceed $1 M_\odot \rm \ yr^{-1}$. The "wind boost" case likely has reached and exceeded a saturation point with MB, and additional effects must be considered to dampen the angular momentum loss for this scheme.
\end{itemize}

The systematic mismatches 
between observed and predicted NS LMXB properties 
seen in previous work such as \cite{Podsiadlowski2002} are again found in our work when the default MB prescription is used. These discrepancies between observations and simulations begin to disappear, however, once we include the effects of convective turnover time and non-isothermal winds. With these changes, the MT rates approach those seen in observed systems, and our simulations more effectively reproduce the samples of persistent, and transient, binaries. With these results in mind, numerical studies of LMXBs that begin their calculations at long periods should no longer use the Skumanich prescription for MB as it does not adequately reproduce observed systems. Instead, the studies need to include additional effects in their MB schemes. 

A clear extension of this work would be to include saturation effects in the MB prescription \citep{Mestel1987}. One source of decreasing the MB strength that is not accounted for in this work, is the change in the magnetic field structure as the period of the system changes. It has been shown that in short period binaries a so-called 'dead zone' is produced, trapping wind material \citep{Mestel1987, Ivanova2003}. This trapped material cannot escape the system, reducing the angular momentum loss through MB. Additionally, the inclusion of irradiation-induced wind and more complex magnetic field structures, such as a dipolar field similar to what is done in \citet{Justham2006}, is possible.

Possibilities for future analysis with the simulations produced for this work include the possibility of determining viable progenitors of observed LMXBs. Using the reproducibility search for observed systems in section \ref{sec:stat}, we can find a rough parameter space that produces progenitors for each of our LMXBs of interest for a given MB prescription. Finding the possible progenitors will act as a "reverse population synthesis" method, where instead of providing initial conditions, we use observed binaries and their progenitors to infer what the initial conditions may have been. The reverse population synthesis method cannot, however, be used with the simple Skumanich law, where many persistent systems are not reproduced.

\section*{Acknowledgements}

We would like the thank the anonymous referee for helpful comments. NI acknowledges support from the CRC program, funding from an NSERC Discovery grant, and acknowledges that a part of this work was performed at the KITP, which is supported in part by the NSF under grant No. NSF PHY17-48958. CH acknowledges support from an NSERC Discovery Grant, and an Accelerator Supplement. This research was enabled by the use of computing resources provided by WestGrid and Compute/Calcul Canada.



\bibliographystyle{mnras}
\bibliography{references} 



\appendix


\bsp	
\label{lastpage}
\end{document}